\documentclass[12pt]{article}
\pdfoutput=1

\usepackage{amsmath}
\usepackage{graphicx}
\usepackage{xcolor}
\usepackage{framed}
\definecolor{shadecolor}{rgb}{0.90,0.90,0.90}
\usepackage{hyperref}
\usepackage{subcaption}
\usepackage[utf8]{inputenc}

\usepackage{setspace}
\usepackage{amsmath, amssymb, amsthm, float, graphicx}
\numberwithin{equation}{section}

\hypersetup{colorlinks=false, linkcolor=blue, citecolor=red}

\def\beq{\begin{eqnarray}}\def\eeq{\end{eqnarray}}
\def\be{\begin{equation}}\def\ee{\end{equation}}
\def\g{\gamma}

\def\m{\mu}

\def\e{\epsilon}
\def\k{\kappa}
\def\b{\beta}
\def\d{\delta}

\def\D{\Delta}
\def\G{\Gamma}
\def\l{\lambda}
\def\pd{\partial}

\def\bz{\bar{z}}

\def\la{\langle}
\def\ra{\rangle}

\def\mo{{\mathcal{O}}}

\def\G{\Gamma}
\def\mc{{\mathcal{C}}}

\usepackage{bm}
\usepackage{bm}

\newmuskip\pFqmuskip

\newcommand*\pFq[6][8]{%
  \begingroup 
  \pFqmuskip=#1mu\relax
  \mathchardef\normalcomma=\mathcode`,
  \mathcode`\,=\string"8000
  \begingroup\lccode`\~=`\,
  \lowercase{\endgroup\let~}\pFqcomma
  {}_{#2}F_{#3}{\left[\genfrac..{0pt}{}{#4}{#5};#6\right]}%
  \endgroup
}
\newcommand{\pFqcomma}{{\normalcomma}\mskip\pFqmuskip}

\newmuskip\pTqmuskip
\newcommand*\pTq[6][8]{%
  \begingroup 
  \pTqmuskip=#1mu\relax
  \mathchardef\normalcomma=\mathcode`,
  \mathcode`\,=\string"8000
  \begingroup\lccode`\~=`\,
  \lowercase{\endgroup\let~}\pTqcomma
  {}_{4}\Theta_{3}{\left[\genfrac..{0pt}{}{#4}{#5}\mid 1,1\right]}%
  \endgroup
}

\newcommand{\pTqcomma}{{\normalcomma}\mskip\pTqmuskip}

\newmuskip\pTTqmuskip
\newcommand*\pTTq[8][6]{%
	\begingroup 
	\pTTqmuskip=#1mu\relax
	\mathchardef\normalcomma=\mathcode`,
	\mathcode`\,=\string"8000
	\begingroup\lccode`\~=`\,
	\lowercase{\endgroup\let~}\pTTqcomma
	{}_{p}\Theta_{q}{\left[\genfrac..{0pt}{}{#4}{#5}\mid {#6},{#7}\right]}%
	\endgroup
}

\newcommand{\pTTqcomma}{{\normalcomma}\mskip\pTTqmuskip}

\begin{document}
\title{\bf Resummation at finite conformal spin}
\date{}

\author{Carlos $\text{Cardona}^{\Delta}$,~$\text{Sunny Guha}^{J}$,~$\text{Surya Kiran Kanumilli}^{J}$ ~and Kallol $\text{Sen}^{\beta}$\\~~~~\\
${}^{\Delta}$Niels Bohr International Academy and Discovery Center,\\University of Copenhagen, Niels Bohr Institute\\ Blegamsvej 17, DK-2100 Copenhagen Ø, Denmark
\\ ~~~~\\
${}^{J}$
George P. and Cynthia Woods Mitchell Institute\\ for Fundamental Physics and Astronomy,\\
Texas A\&M University,
\\ College Station, TX 77843, USA
\\~~~~\\
${}^{\beta}$Kavli Institute for the Physics and Mathematics of the Universe (WPI),\\
 University of Tokyo, Kashiwa, Chiba 277-8583, Japan}

\maketitle
\vskip 2cm
\abstract{We generalize the computation of anomalous dimension and correction to OPE coefficients at finite conformal spin considered recently in  \cite{Cardona:2018dov, Liu:2018jhs} to arbitrary space-time dimensions. By using the inversion formula of Caron-Huot and the integral (Mellin) representation of conformal blocks, we show that the contribution from individual exchanges to anomalous dimensions and corrections to the OPE coefficients for ``double-twist" operators $[\mo_1\mo_2]_{\D,J}$ in $s-$channel can be written at finite conformal spin in terms of generalized Wilson polynomials. This approach is democratic with respect to space-time dimensions, thus generalizing the earlier findings to cases where closed form expressions of the conformal blocks are not available. }

\vfill {\footnotesize 
	carlosgiraldo@nbi.ku.dk,\ \ \ sunnyguha@physics.tamu.edu, \ \ \ suryak@physics.tamu.edu \ \ and \ \ kallol.sen@ipmu.jp}

\newpage

\tableofcontents

\onehalfspacing

\newpage

\section{Introduction}
The Conformal Bootstrap program has proven to be very successful in recent years because of the important constraints imposed on generic Conformal Field Theories that can be extracted by both numerical \cite{Rattazzi:2008pe, El-Showk:2014dwa, Kos:2014bka}\footnote{See recent reviews \cite{Rychkov:2016iqz,Simmons-Duffin:2016gjk,Poland:2018epd}} and analytical  \cite{Komargodski:2012ek, Fitzpatrick:2012yx, Simmons-Duffin:2016wlq} techniques in any dimension and at any coupling. The lightcone limit of crossing equation for a four-point function provides us with a particular amenable analytical region that contains important physical information. This limit is controlled by large spin operators which allows one to develop a systematic perturbative expansion of the crossing relation in terms of inverse spin \cite{Alday:2015eya, Alday:2015ota,Alday:2015ewa,Kaviraj:2015cxa,Kaviraj:2015xsa}.

Recently, a formula that inverts the partial-wave expansion of a four-point function has been developed \cite{Caron-Huot:2017vep, Simmons-Duffin:2017nub}, which in particular can be used to resum the expansion in large spin, providing access to anomalous dimension and OPE coefficients at finite values of the conformal spin, as has been done recently in four dimensions \cite{Cardona:2018nnk, Liu:2018jhs, Sleight:2018epi}. Previously, an analogous expansion for the large spin was computed in the series of works \cite{Alday:2016njk} and applied to holographic CFTs in four dimensions, large $N-$theories in three dimensions and for $\mathcal{N}=4$ SYM. Some expressions in arbitrary dimensions were also given in \cite{Cardona:2018nnk}, which even though resumming the large sum expansion, are only valid asymptotically. The reason is that in \cite{Cardona:2018nnk} the contribution coming from the residues in Mellin space which were subleading in large $\beta$ (conformal spin) were neglected.

In this paper we follow up on the computation of anomalous dimensions and OPE corrections for double-twist operators from the inversion formula \cite{Caron-Huot:2017vep,  Simmons-Duffin:2017nub} initiated in \cite{Cardona:2018nnk} by including all the sub-leading residues which results in an analytically continued expression valid at any value of the conformal spin and in arbitrary dimension.

We will consider the correlation function of four conformal primary scalar operators given by conformal invariance as,
\be \label{4p}
\langle {\cal O}_4(x_4)\cdots{\cal O}_1(x_1)\rangle = 
\frac{1}{(x_{12}^2)^{\frac12(\Delta_1+\Delta_2)}(x_{34}^2)^{\frac12(\Delta_3+\Delta_4)}}
\left(\frac{x_{14}^2}{x_{24}^2}\right)^{a}
\left(\frac{x_{14}^2}{x_{13}^2}\right)^{b}
{\cal G}(z,\bz)
\ee
where $a=\frac12(\Delta_2-\Delta_1)$, $b=\frac12(\Delta_{3}-\Delta_4)$,
and $z$, $\bz$ are conformal cross-ratios given by,
\be
u=z\bz=\frac{x_{12}^2x_{34}^2}{x_{13}^2x_{24}^2},\qquad v=(1-z)(1-\bz) = \frac{x_{23}^2x_{14}^2}{x_{13}^2x_{24}^2}\,.
\ee
Henceforth, we will be using $(z,\bz)$ coordinates instead of $(u,v)$. The function ${\cal G}(z,\bz)$ has the following {\it s-channel} conformal block expansion representation,
\be \label{OPE}
{\cal G}(z,\bz) = \sum_{J,\Delta} f_{12{\cal O}}f_{43{\cal O}}\,G_{\Delta,J}(z,\bar{z})\, , 
\ee
where the sum runs over the exchanged primary operators with spin $J$ and dimension $\Delta$.
$G_{\Delta,J}$ are the conformal blocks eigenfunctions of the quadratic and quartic Casimir invariants of the conformal group and which can be conveniently represented by  the following spectral representation \cite{Costa:2012cb},
\be\label{spectralexpansion}
{\cal G}(z,\bz) = 1+
\sum_{J=0}^\infty \int_{d/2-i\infty}^{d/2+i\infty} \frac{d\Delta}{2\pi i} \,C(\Delta,J)\,f_{\Delta,J}(z,\bz)\,. 
\ee
The function $f_{\Delta,J}$ is given in terms of a linear combination of conformal blocks plus its shadow respectively as,
\be\label{lincomb}
f_{\D,J}(u,v)=\frac{1}{k_{d-\D,J}}\frac{\g_{\l_1,a}}{\g_{\bar{\l}_1,b}}G_{\D,J}(u,v)+\frac{1}{k_{\D,J}}\frac{\g_{\bar{\l}_1,a}}{\g_{\l_1,a}}G_{d-\D,J}(u,v)\,,
\ee
with coefficients defined in appendix \ref{intrep}. The appropriate normalization for the integral representation, to match with the physical conformal block is given in \eqref{intr1} of appendix \ref{intrep}\footnote{Note that $\g_{a,b}$ used in the normalization is different from the $\g^{\D,J}_{12}$ used for the notation of the anomalous dimension. We have used the notations of \cite{Dolan:2011dv}.}. For each operator exchange, labelled by $(\D,J)$, the contour integral representation of $f_{\D,J}(u,v)$ given in \eqref{intr}, picks up the physical and shadow poles to give the linear combination on the {\it rhs} of \eqref{lincomb}.

Our main tool in this work is the Lorentzian OPE inversion formula \cite{Caron-Huot:2017vep,Simmons-Duffin:2017nub}, which allows us to extract $C(\Delta,J)$ from the discontinuities of the four-point function.
\be\label{exactC}
C^t(\Delta,J) = \frac{\kappa_{J+\Delta}}{4}\int_{0}^1 dz d\bz\, \mu(z,\bz)\, G_{J+d-1,\Delta+1-d}(z,\bz)\,{\rm dDisc}\big[{\cal G}(z,\bz)\big]\,,
\ee
where the conformal invariant measure is given by,
\be\label{measure}
\mu(z,\bz) = \left|\frac{z-\bz}{z\bz}\right|^{d-2} \frac{\big((1-z)(1-\bz)\big)^{a+b}}{(z\bz)^2}\,. 
\ee
The partial wave coefficient is given as,
\be
C(\Delta,J)=C^t(\Delta,J)+(-1)^J C^u(\Delta,J)\,,
\ee
The $u$-channel contribution $C^u$ is computed from the same integral \eqref{exactC} but with 1 and 2 interchanged and
the integration ranging from $-\infty$ to 0 and the double discontinuity taken around $z=\infty$.  
In practice, the OPE coefficients can be extracted from the $\bz$ integration as a power expansion in small $z$. At leading order in small $z$ \eqref{exactC} is approximated by,
\be
C^t(\Delta,J)=\int_0^1 \frac{dz}{2z}z^\frac{\tau}{2}\,C^t(z,\beta)\,,
\ee
where the following ``generating function'' has been defined,
\be\label{generating}
C^t(z,\beta)\equiv\int_z^1{d\bz\,(1-\bz)^{a+b}\over \bz^2}\kappa_{\beta}\,k_{\beta}(\bz)\text{dDisc}[\mathcal{G}(z,\bz)]\,,
\ee
with
\begin{equation}\label{kappa}
	k_{2h}(z)=z^h\pFq{2}{1}{h,h}{2h}{z}.
\end{equation}
The usual conformal twist and spin are respectively $\tau=\D-J$ and $\b=\D+J$.
We are interested in studying the contributions to \eqref{generating}  coming from a single exchange, so by using the $t-$channel block decomposition of the four-point point function $\mathcal{G}(z,\bz)$ we can compute the contribution:
\beq\label{genfunc}
C^t(z,\b)|_{\Delta,J}&=&f_{14(\D,J)}f_{23(\D,J)}\k_\b\int_z^1d\bz\ \frac{(1-\bz)^{a+b}}{\bz^2}k_\b(\bz)\text{dDisc}\bigg[\frac{(z\bz)^\frac{\D_3+\D_4}{2}G_{\D,J}(1-z,1-\bz)}{[(1-z)(1-\bz)]^\frac{\D_2+\D_3}{2}}\bigg]\, ,\nonumber\\
\eeq
where $f_{i\,j(\D,J)}$ corresponds to the OPE structure constant between the external scalars $i$ and $j$ and the exchanged operator. 

At small $z$  the generating function \eqref{genfunc} can be written as a power expansion in $z$, whose contribution at the leading term from a single exchange will be given by
\be
C^t(z,\b)|_{\Delta,J}\sim C(\b) z^{{\tau\over2}+{1\over 2}\gamma_{12}(\beta)}\,,
\ee
where $C(\b)$ and  $\gamma_{12}(\beta)$ corresponds to the square OPE coefficient and anomalous dimension of the double twist operator having $\tau=-(\D_1+\D_2)$.
If the anomalous dimension $\gamma_{12}(\beta)$ and correction to OPE coefficients  $\delta P_{\D,J}(\b)$ are small, we can write
\be
C(\beta) = C_0(\beta)[1+\delta P_{\D,J}(\b)]\,, 
\ee
so that,
\be\label{small_gen}
 C^t(z,\b)|_{\Delta,J}\sim z^{{\tau\over2}} C_0(\b) \bigg(\delta P_{\D,J}(\b) +{1\over 2}\gamma_{12}(\beta)\log(z)\bigg)\,.
\ee
We similarly need to expand the RHS of \eqref{genfunc} at small $z$, where the conformal blocks develop log-terms and regular terms, as reviewed on the Appendix. Therefore  we can see that the anomalous dimension will be related to the log terms, whereas the OPE coefficients will be given by the regular terms. In this paper we will restrict to the four point function of identical scalars $\phi$ ($\D_1=\D_2=\D_3=\D_4=\D_\phi$). We focus on the anomalous dimensions and corrections to the OPE coefficients for double twist operators of the form $[\phi\phi]_J=\phi\pd_{\m_1}\dots\pd_{\m_J}\phi$.  

 \section{Warming up}\label{sec1}
In four and two dimensions the conformal blocks can be represented by combinations of Gauss hypergeometric functions through \eqref{kappa}. Respectively we have, 
\begin{flalign}\label{blocks24}
	&& G_{\Delta,J}(z,\bz) &= \frac{k_{\Delta-J}(z)k_{\Delta+J}(\bz)+k_{\Delta+J}(z)k_{\Delta-J}(\bz)}{1+\delta_{J,0}}
	\,,\quad\text{2D}
	\\
	&& G_{\Delta,J}(z,\bz) &= \frac{z\bz}{\bz-z}\big[
	k_{\Delta-J-2}(z)k_{\Delta+J}(\bz)-k_{\Delta+J}(z)k_{\Delta-J-2}(\bz)\big]\,,\quad\text{4D}\,. & 
\end{flalign}
Hence in the small$-z$ limit, the building-block integral we need to perform is of the form,
\begin{equation}\label{3.30}
	J_0\equiv\int_0^1 \frac{dz}{z^2}\big(\frac{z}{1-z}\big)^pk_h(z)k_g(1-z).
\end{equation}
This integral is a special case of a Jacobi transform, which has been studied in detail recently in the context of one dimensional Conformal Field Theories in \cite{Hogervorst:2017sfd}\footnote{In the lightcone limit, the conformal blocks factorise and the kernel for the inversion formula can be written in terms of one dimensional integrals as in \eqref{3.30}}. \eqref{3.30} computes the crossing kernel in the lightcone limit even in higher dimensions, because of the factorization property of the blocks as we see from \eqref{blocks24}.
This type of integrals are hard to perform in position space, but as we are going to see, they are straightforward in Mellin space. We will evaluate this simple example in detail as it captures all the conceptual details involved in the more complicated integrals dealt later in the text.

We follow the same strategy as in \cite{Hogervorst:2017sfd}. First we will expand both $k_h(z)$ functions in the more convenient variable $\frac{z}{1-z}$, by using the following identity of the hypergeometric functions,
\begin{equation}\label{identity}
	{}_2F_1(h,h,2h,z)=(1-z)^h{}_2F_1(h,h,2h,\frac{z}{z-1}) .
\end{equation}
Then representing the hypergeometrics using the Mellin-Barnes representation we will be able to perform the $z$ integral first.
The Mellin-Barnes form of hypergeometric is given by,
\begin{equation}\label{mellinbarne}
	{}_2F_1(a,b;c;z) =\frac{\Gamma(c)}{\Gamma(a)\Gamma(b)} \int_{-i\infty}^{i\infty} \frac{\Gamma(a+s)\Gamma(b+s)\Gamma(-s)}{\Gamma(c+s)}(-z)^s\,ds.
\end{equation}
Using $\ref{identity}$ and $\ref{mellinbarne}$ , the integral $\ref{3.30}$ becomes,
\begin{equation}\label{j0}
	J_0=\frac{\Gamma(2h)\Gamma(2g)}{\Gamma(h)^2\Gamma(g)^2}\int_\mc ds \int_\mc dt \int_0^1 dz \frac{z^{p+s-t+h-g-2}}{(1-z)^{p+s-t+h-g}} \frac{\Gamma(-s)\Gamma(h+s)^2\Gamma(-t)\Gamma(t+g)^2}{\Gamma(2h+s)\Gamma(2g+t)} 
\end{equation}
$\mathcal{C}$ refers to the contour going from $-i\infty$ to $+i\infty$ and encircling the right half of the plane. The contribution of the semi-circular arc at $\infty$ vanishes. The $z$-integral is log-divergent, so we need to regularize it. To perform it we follow the prescription of \cite{Hogervorst:2017sfd} and deform one of the hypergeometrics in the following way,
\begin{equation}
	k_h(z)=z^hz^{\epsilon}\pFq{2}{1}{h,h}{2h+\epsilon}{z}
\end{equation}
Now the $z$-integral becomes a simple beta function and the Mellin integration over $s$ can be performed by means of the Barnes' second lemma:
\begin{align}\label{secondlemma}
	\nonumber
	 \int_{-i\infty}^{i\infty} & \frac{\Gamma(a+s) \Gamma(b+s)\Gamma(c+s)\Gamma(1-d-s)\Gamma(-s)}{\Gamma(e+s)}ds= \\ &\frac{\Gamma(a)\Gamma(b)\Gamma(c)\Gamma(1-d+a)\Gamma(1-d+b)\Gamma(1-d+c)}{\Gamma(e-a)\Gamma(e-b)\Gamma(e-c)}
\end{align}
where we  should take,
\begin{equation}
	a=h \quad b=h \quad c=-1+p-t+h-g+\epsilon \quad d=p-t+h-g \quad e=2h+\epsilon
\end{equation}
This gives us the following result:
\begin{equation}
	\frac{\Gamma(2h+\epsilon)\Gamma(2g)}{\Gamma(g)^2}\int_{-i\infty}^{i\infty}\, dt\,\frac{\Gamma(-1+p-t+h-g+\epsilon)\Gamma(1-p+t+g)\Gamma(1-p+t+g)\Gamma(-t)\Gamma(g+t)^2}{\Gamma(h+\epsilon)^2\Gamma(2h+1-p+t-h+g)\Gamma(2g+t)} .
\end{equation}
Notice that the divergence in $1/\epsilon$ automatically cancels and now we can safely take the $\epsilon \to 0$ limit. What remains is doing the contour over the t-variable. By closing the  contour to the right, there are two sets of poles for the t-variable,
\begin{equation}\label{poleex}
	t\in\mathbb{N} \qquad t\in-1+p+h-g+\mathbb{N}
\end{equation}

Summing up these two series of residues, we get the following result,
\begin{align}\label{final}
	\nonumber
	J_0=&\frac{\Gamma(2h)\Gamma(1+g-p)^2\Gamma(-1+h-g+p)}{\Gamma(h)^2\Gamma(1+h+g-p)} \pFq{4}{3}{g,g,1+g-p,1+g-p}{2g,2-h+g-p,1+h+g-p}{1} \\
	& +\frac{\Gamma(2g)\Gamma(1-h+g-p)\Gamma(-1+h+p)^2}{\Gamma(g)^2\Gamma(-1+h+g+p)}\pFq{4}{3}{h,h,-1+h+p,-1+h+p}{2 h,h-g+p,-1+h+g+p}{1}
\end{align}
An observation from this example which we will apply to the remaining cases considered below is in order. Naively, we could have started by trying to compute the integral \eqref{3.30} by using the usual series expansion of the hypergeometric function. However, since this is only convergent in the region $|z|<1$, this will  produce an asymptotic expansion valid only for large values of $h$, as that is the regime controlled by the small $z$ region. Continuing to finite $h$ involves re-suming additional contributions from the lower limit of the $z$ integral\footnote{The lower limit of the $z$ integral is not convergent and gives rise to additional contributions discussed in \cite{Liu:2018jhs}.}.    The Mellin-Barnes form in \eqref{mellinbarne}, makes these additional contributions explicit, in terms of the second pair of poles in \eqref{poleex}.

\section{Anomalous dimension}
In this section we calculate the contribution to anomalous dimension of a double-twist operator from a single block exchange of a four-point correlation function of identical operators, by using the integral representation of conformal block. So for our case $\tau=-(\D_1+\D_2)=-2\D_\phi$ and the conformal spin $\b=\D+J$ defined for the double twist operators in the $s-$channel. The anomalous dimensions $\g_{12}^{\D,J}(\b)$ are the corrections to the dimensions of operators $[\phi\phi]_J\equiv \phi\pd_{\m_1}\dots\pd_{\m_J}\phi$, given by,
\be
\D_{[\phi\phi]_J}=2\D_\phi+J+1/2\g_{12}^{\D,J}\,,
\ee
due to exchange of operators of dimension $\D$ and spin $J$ in the crossed ($t$) channel. We restrict ourselves to corrections to the double twist operators $\sim z^{\frac{\D_1+\D_2}{2}}\log z$ which comes only from the leading $\log z$ term {\it i.e} the leading twist contributions in the crossed ($t$) channel. Please note that we refer to $\tau$ for the double twist operators and not the twist of the $t-$channel exchanges. Also for clarification, we move back and forth between the notations $d/2$ and $h=d/2$ in what follows. 

\subsection{Scalar exchange}

When the anomalous dimension is small, contribution from the exchange can be computed as \cite{Caron-Huot:2017vep}:
\begin{equation}\label{invscalar}
	\gamma_{12}^{\Delta,0}(\beta)=\frac{1}{C_0(\b)} \int_0^1 \frac{d\bar{z}}{\bar{z}^2} \kappa_{\beta} k_{\beta}(\bar{z}) \mathrm{dDisc}\left[ \left(\frac{1-\bar{z}}{\bar{z}}\right)^\frac{\tau}{2}G^t_{\Delta,0}\vert_{\log} \right].
\end{equation}
In the above equation, $G^t_{\Delta,0}\vert_{\log}$ stands for the log term in the $z\rightarrow 0$ expansion of t-channel conformal block. First contribution to the OPE coefficient comes from unity block and is given by \cite{Heemskerk:2009pn, Komargodski:2012ek, Fitzpatrick:2012yx}   
\begin{equation}
	C_0(\b) = \frac{\Gamma \left(\frac{\beta }{2}\right)^2 \Gamma \left(\frac{1}{2} (\beta -\tau -2)\right)}{\Gamma (\beta -1) \Gamma \left(-\frac{\tau }{2}\right)^2 \Gamma
		\left(\frac{1}{2} (\beta +\tau +2)\right)}\,.
\end{equation} 
In the second part we will be dealing with the corrections to these coefficients from conformal bock exchanges.
The constant $\kappa_{\beta}$ is,
\begin{equation}
		\kappa_\beta = \frac{\Gamma \left(\frac{\beta }{2}\right)^4}{2 \pi ^2 \Gamma (\beta -1) \Gamma (\beta )}.
\end{equation}
Our starting point will be the Mellin transform integral representation for the scalar conformal block in the t-channel detailed in the Appendix \ref{intrep}, 
\begin{equation}\label{log_scal_block_mell}
	G^t_{\Delta,0}\vert_{\log} = - \frac{\Gamma (\Delta )  \Gamma (-h+\Delta +1)}{ \Gamma \left(\frac{\Delta }{2}\right)^3 \Gamma \left(-h+\frac{\Delta }{2}+1\right)} \int_\mc ds \left( \frac{1-\bar{z}}{\bar{z}}\right)^{ \frac{\Delta}{2}+s}    \Gamma(-s) \frac{\Gamma(s+\frac{\Delta}{2}) \Gamma(1-h+s+\frac{\Delta}{2})}{\Gamma(1-h+\Delta+s)} ,
\end{equation}
where and we have included the global factor from the crossing equation.
We can immediately check that by picking up $s=n$ poles for the s-integral and summing over residues, we get the well known log-term for the scalar block \cite{Dolan:2000ut,Dolan:2011dv}, 
\begin{equation}\label{log_scal_block}
	- \frac{(1-\bar{z})^{\frac{\Delta}{2}+\tau /2}}{\bar{z}^{\frac{\Delta}{2}+\tau /2}}\frac{\Gamma(\Delta)}{\Gamma(\frac{\Delta}{2})^2}\:{}_2F_1(1-\frac{d}{2}+\frac{\Delta}{2},\frac{\Delta}{2},1-\frac{d}{2}+\Delta,\frac{\bar{z}-1}{\bar{z}}) \,,
\end{equation}
thus confirming that \eqref{log_scal_block_mell} is the correct representation to be used.

Taking the double discontinuity over expression \eqref{log_scal_block_mell} produces the following additional phase
\begin{equation}
	2 \sin ^2\left[ \pi  \left(\lambda_2+\frac{\tau}{2} +s\right)\right] \,.      
\end{equation}
Note however that in the inversion formula \eqref{generating}, we are considering the discontinuities of individual physical blocks in $t-$channel. These physical blocks are reproduced by the $s=n$ poles in \eqref{log_scal_block_mell} for $n\in\mathbb{I}_{\geq0}$ respectively. Hence the phase factor corresponding to the entire block simply multiplies the whole integral by  $\sin ^2\left[ \pi \left( \lambda_2+ \frac{\tau}{2} \right)   \right] $.
Considering this phase and plugging in the representation \eqref{log_scal_block} into \eqref{invscalar}, the contribution to the anomalous dimension coming from the scalar exchange is then  given by,
\begin{align}
	\gamma^{\Delta,0} = &-4 \sin ^2\left[ \pi  \left(\frac{\Delta +\tau}{2} \right)\right] \frac{\kappa_\beta}{C_0(\b)} \frac{ \Gamma (\beta ) \Gamma (\Delta )  \Gamma (-h+\Delta +1)}{\Gamma
		\left(\frac{\beta }{2}\right)^2 \Gamma \left(\frac{\Delta }{2}\right)^3 \Gamma \left(-h+\frac{\Delta }{2}+1\right)} \nonumber \\ &\int_\mc ds\, dt  \int_0^1 \frac{d\bar{z}}{\bar{z}^2}  \left( \frac{\bar{z}}{1-\bar{z}}\right)^{ \frac{\beta -\Delta -\tau}{2}   -s+t} \Gamma(-t)\frac{\Gamma\left( \frac{\beta}{2} + t\right)^2 }{\Gamma\left( \beta + t\right)} \Gamma(-s) \frac{\Gamma(s+\frac{\Delta}{2}) \Gamma(1-h+s+\frac{\Delta}{2})}{\Gamma(1-h+\Delta+s)}. \label{an1}
\end{align}
This is essentially the same integral \eqref{j0} that we have dealt with in section \ref{sec1} and hence the same steps follows, leading us to,
\begin{align}\label{anomscalarmaster}
	\gamma^{\Delta,0} =-4 &\frac{\kappa_{\beta} \, \Gamma (\Delta )\sin ^2\left(\frac{\Delta +\tau}{2} \pi  \right)}{\pi ^2 \Gamma\left(\frac{\Delta }{2}\right)^3 C_0(\beta)} \nonumber \\
	&\times \Bigg( \frac{\Gamma (\beta ) \Gamma \left(\frac{\Delta }{2}\right) \Gamma \left(\frac{\Delta +\tau +2}{2}\right)^2 \Gamma \left(\frac{\beta -\Delta -\tau -2}{2} \right)}{\Gamma \left(\frac{\beta }{2}\right)^2 \Gamma \left(\frac{\beta +\Delta +\tau +2}{2}\right)}    \, _4F_3\bigg[\begin{matrix}1-h+\frac{\Delta }{2},\frac{\Delta }{2},\frac{\Delta + \tau}{2}+1,\frac{\Delta + \tau}{2}+1\\ 1-h+\Delta,2-\frac{\beta - \Delta - \tau
		}{2},\frac{\beta + \Delta + \tau}{2}+1\end{matrix};1\bigg] \nonumber \\ 
	&+\frac{\Gamma \left(\frac{\beta -\tau -2}{2}\right) \Gamma (-h+\Delta +1) \Gamma \left(\frac{-\beta +\Delta +\tau +2}{2}\right) \Gamma \left(\frac{-2 h+\beta -\tau }{2} \right)}{\Gamma \left(\frac{-2 h+\Delta +2}{2}\right) \Gamma \left(\frac{-2 h+\beta +\Delta -\tau }{2} \right)} \, _4F_3\bigg[\begin{matrix}\frac{\beta }{2},\frac{\beta }{2},\frac{\beta -\tau}{2}-1,\frac{\beta -\tau}{2}-h\\ \beta,\frac{\beta-\Delta-\tau }{2},-h+\frac{\beta + \Delta - \tau }{2}\end{matrix};1\bigg] \Bigg).
\end{align}

The first Hypergeometric comes from summing over residues at $s=n$ pole series and second from those at $s = \frac{\beta-\Delta-\tau}{2} -1 +n$. In four dimensions (i.e. $h=2$), this expression matches with (3.56) of \cite{Liu:2018jhs}. Following \cite{Hogervorst:2017sfd}, we can write ($\ref{anomscalarmaster}$) in terms of more compact Wilson function $\phi_{\alpha}$ \cite{2003math......6424G} as,
\begin{align}
\gamma^{\Delta,0} = -4&\frac{\kappa_{\beta}}{C_0(\beta)}\frac{\sin ^2\left((\frac{\Delta +\tau}{2}) \pi \right)}{\Gamma\left(\frac{\Delta}{2}\right)^2} \,\Gamma (\b) \Gamma (\Delta )  \Gamma (1-h+\Delta ) \Gamma \left(\frac{\beta -\tau}{2} - 1 \right) \nonumber \\\times  &  \Gamma\left( \frac{\beta-\tau}{2}-h \right) \Gamma\left( 1+\frac{\Delta+\tau}{2}\right)^2 
\phi_{\alpha} \left(\beta;a,b,c,d\right)   \,,
\end{align}
where
\begin{align}\label{workingsset}
\begin{split}
a&=\frac{\Delta}{2}+\frac{1}{2}+\frac{\tau}{4}, \hspace{2cm} b=\frac{1}{2} +\frac{\Delta}{2}-\frac{\tau}{4}-h\,, \hspace{2cm} c=\frac{\beta}{2}+\frac{\tau}{4}+\frac{1}{2}\,,\\
d&=\frac{\beta}{2}-\frac{\tau}{4}-\frac{1}{2}\,, \hspace{2cm} \alpha=\frac{h}{2}+\frac{\tau}{4}\,, \hspace{3.4cm}\beta=\frac{1}{2}+\frac{\tau}{4}\,.
\end{split}
\end{align}
We can rewrite the scalar contribution to the anomalous dimension as a ${}_7F_6$ hypergeometric: 
\footnotesize{
	\begin{align}
	\gamma^{\Delta,0} = -&4 \frac{\kappa_{\beta}}{C_0(\beta)}
	\frac{\sin ^2\left[(\frac{\Delta +\tau}{2}) \pi  \right] \Gamma (\Delta ) \Gamma \left(\frac{\beta -\tau}{2} -1 \right)  \Gamma\left(\beta  \right) \Gamma\left(\frac{\beta -\tau}{2} -h\right) \Gamma\left( 1+\frac{\Delta+\tau}{2}\right)^2 \Gamma \left(-h+\frac{\beta }{2}+\Delta +1\right)}{\Gamma(\frac{\beta}{2}) \Gamma\left(\frac{\Delta }{2}\right)^2 \Gamma \left(\frac{\beta +\Delta }{2}\right) \Gamma \left(\frac{\beta +\Delta
			+\tau}{2} +1\right) \Gamma \left(1-h+\frac{\beta +\Delta}{2} \right) \Gamma \left(\frac{\beta
			+\Delta -\tau}{2} -h\right)}   \nonumber \\ &_7F_6\bigg[\begin{matrix}\frac{\beta }{2},1-h+\frac{\Delta }{2},1-\frac{h}{2}+\frac{\beta}{4}+\frac{\Delta}{2},\frac{\Delta}{2},-h+\frac{\beta}{2}+\Delta,-h+\frac{\Delta-\tau}{2},1+\frac{\Delta+\tau}{2}\\ -\frac{h}{2}+\frac{\beta}{4}+\frac{\Delta}{2},\frac{\beta+\Delta}{2},1-h+\frac{\beta+\Delta}{2},1-h+\Delta,-h+\frac{\beta + \Delta - \tau }{2},1+\frac{\beta + \Delta + \tau }{2}\end{matrix};1\bigg].
	\end{align}
}
The same kind of ${}_7F_6$ function also appears in the context of the $\e-$expansion, which is recently discussed in \cite{Gopakumar:2018xqi}. 

\subsection{Spin exchange}
	The computation of the contribution to the anomalous dimension from a scalar can be straightforwardly upgraded to the spinning exchange, which is the topic of this section.  We start with the integral representation of the log term for the conformal block, as discussed in  the Appendix \ref{intrep}, 
\begin{align}
		\begin{split}\label{spin_block_mell}
			G_{\D,J}(1-z,1-\bz)=&-\frac{\G(2\l_1)}{\G(\l_1)^2(d-2)_J}\log z\sum_{m=0}^J (-1)^m \frac{A_m(J,\D)}{(\l_1-m)_m^2}\\
			&\times\int_\mc ds\ \bigg(\frac{1-\bz}{\bz}\bigg)^{s+\l_1-m}\G(-s)\frac{(\l_1-m)_s(1-\bar{\l}_2)_s}{(1+\l_2-\bar{\l}_2)_{s+J-m}}\,.
		\end{split}
	\end{align}
Using the spin block \eqref{spin_block_mell} instead of the scalar block  in \eqref{invscalar} and evaluating the sum over residues 
at the poles $s=n$ and $s = \frac{\beta-\Delta-J-\tau}{2} +m -1 +n$  we get 
	
	\begin{align}\label{anomspinmaster}
		\gamma^{\Delta,J}& =4\frac{\kappa_{\beta}}{C_0(\beta)} \sum_{m=0}^J \frac{(-1)^{m+1}}{(d-2)_J} \frac{ A_m(J,\Delta)}{(\lambda_1 -m)^2_m}         
		\frac{\Gamma (-h+\Delta +1) \Gamma (2 \l_1 )}{\Gamma \left(\l_1\right)^2 \Gamma \left(\l_1-m\right)}\sin ^2\left(\frac{1}{2} \pi  (\Delta +J+\tau )\right) \nonumber \\
		&\Bigg( \frac{\Gamma (\beta ) \Gamma \left(\l_1-m \right) \Gamma \left(\frac{\tau +2}{2} +\l_1-m\right)^2 \Gamma \left(\frac{\beta -\tau -2}{2} -\l_1+m\right)}{\Gamma \left(\frac{\beta }{2}\right)^2 \Gamma (-h-m+2\l_1 +1) \Gamma \left(\frac{\beta  +\tau +2}{2} +\l_1-m\right)}   \nonumber \\
			& \times _4F_3\bigg[\begin{matrix}1-h+\l_1,\l_1-m,\frac{\tau}{2}+\l_1-m+1,\frac{\tau}{2}+\l_1-m+1\\ -h+2\l_1-m+1,\frac{-\beta+\tau}{2}+\l_1-m+2,\frac{\beta+\tau}{2}+\l_1-m+1\end{matrix};1\bigg]  \nonumber \\		
		&+\frac{\Gamma \left(\frac{\beta -\tau -2}{2} \right) \Gamma \left(\frac{-2 h+2 m+\beta -\tau }{2} \right) \Gamma \left(\frac{2-\beta +\tau}{2} +\l_1-m \right)}{\Gamma \left(\frac{-2 h +2}{2} +\l_1 \right) \Gamma \left(\frac{\beta-2 h -\tau }{2} +\l_1\right)} \, _4F_3\bigg[\begin{matrix}\frac{\beta }{2},\frac{\beta }{2},\frac{\beta -\tau}{2}-1,\frac{\beta -\tau}{2}-h+m\\ \beta,\frac{\beta-\tau }{2}-\l_1+m,-h+\frac{\beta  - \tau }{2}+\l_1\end{matrix};1\bigg] \Bigg)\,.
	\end{align}
	Which as in the scalar case can be written more compactly in terms of a Wilson function as,

\begin{align}
&\gamma^{\Delta,J} = 4\frac{\kappa_\beta}{C_0(\beta)}\sum_{m=0}^J \frac{(-1)^{m+1}}{(d-2)_J} \frac{ A_m(J,\Delta)}{(\lambda_1 -m)^2_m} \frac{\Gamma (\beta ) \Gamma (2 \, \l_1 ) \Gamma \left(\frac{\beta -\tau}{2} -1\right) \Gamma \left(\frac{\beta -\tau}{2} -h+m\right) \Gamma
	\left(\frac{\tau}{2} +\l_1 -m + 1\right)^2 }{ \Gamma\left(\frac{\b}{2}\right) \Gamma \left(\l_1 \right)^2 \Gamma
	\left(\frac{\beta}{2}+\l_1 +1-h\right)  \Gamma \left(\frac{\beta}{2}+\l_1 -m\right)} \nonumber \\ 
&\frac{\sin^2 \left(\pi  \left(\frac{\tau }{2}+\l_1 -m \right)\right) \Gamma \left(-h+2\l_1-m+\frac{\beta }{2}+1\right)\Gamma (-h+\Delta +1) }{\Gamma \left(\frac{\beta -\tau}{2}-
	h+ \l_1  )\right) \Gamma \left(\frac{\beta +\tau}{2} +\l_1 - m +1)\right) \Gamma (-h-m+2\l_1 +1)} \nonumber \\
&_7F_6\bigg[\begin{matrix}\frac{\beta }{2},\l_1-h+1,\l_1-m,\frac{\beta }{4}+\l_1-\frac{h+m}{2}+1,\frac{\beta }{2}+2\l_1-h-m,\l_1-h-\frac{\tau }{2},\l_1-m+\frac{\tau }{2}+1\\ \frac{\beta}{4}+\l_1-\frac{h+m}{2},\frac{\beta}{2}+\l_1-h+1,\frac{\beta }{2}+\l_1-m,2\l_1 -h-m+1,\frac{\beta-\tau }{2}+\l_1-h,\frac{\beta+\tau}{2}+\l_1-m+1\end{matrix};1\bigg].
\end{align}	
	
\section{Corrections to OPE coefficients}\label{opecorrec}
		Similarly as the contribution to the anomalous dimension at leading order in the light-cone limit is given by the log$(z)$ factors, the OPE coefficient corrections corresponding to the given exchanges can be computed by performing the same exercise on the remaining non-log terms coming from the double poles at $t=0$ in the integral representation of the conformal blocks. In the following we shall compute such contributions. Corrections to the OPE have the following form,
		\begin{equation}\label{OPEcorrectionmaster}
			\delta P_{\D,J} =	\frac{1}{C_0(\beta)} \int_0^1 \frac{d\bar{z}}{\bar{z}^2} \kappa_\beta k_{\beta}(\bar{z}) \mathrm{dDisc}\left[ \left(\frac{1-\bar{z}}{\bar{z}}\right)^\frac{\tau}{2} G^t_{\Delta,J}\vert_{\text{reg}} \right].  
		\end{equation}
		We will consider a general spin-$J$ exchange case. Our starting point would be the integral representation of the regular terms of the conformal block calculated in ($\ref{opechannel}$). For both the anomalous dimension ($\ref{invscalar}$) and OPE correction ($\ref{OPEcorrectionmaster}$), there are three sets of integrals : a) the $\bar{z}$ integral, b) kernel integral in Mellin Barnes, and c) the integral coming from the t-channel conformal block representation ($\ref{llog}$,$\ref{opechannel}$).To maintain uniformity we will transform the kernel in $\frac{\bar{z}-1}{\bar{z}}$ variable using,
\begin{equation}\label{identity}
	{}_2F_1(h,h,2h,z)=(1-z)^h{}_2F_1\bigg[h,h,2h,\frac{z}{z-1}\bigg] .
\end{equation}
After this transformation we will write the kernel hypergeometric as a Mellin-Barnes integral ($\ref{mellinbarne}$) in the $t$-variable. Now we are ready to perform the entire $\bar{z}$ integral involving spins. Collecting all powers of $\bar{z}$ from the kernel and the conformal block  the $\bar{z}$ integral becomes,
\begin{equation}
\int_0^1 \frac{d\bar{z}}{\bar{z}^2} \bar{z}^\epsilon \left( \frac{\bar{z}}{1-\bar{z}}\right)^{ \frac{\beta -\Delta-J -\tau}{2} +m  -s+t}  = \frac{\Gamma(\alpha+\epsilon+t)\Gamma(-\alpha-t)}{\Gamma(\epsilon)} \, ,
\end{equation}
with $\alpha = \frac{\beta-\tau}{2}-\l_1 +m -1 -s$. The gamma functions contain terms that are mixed in $s$ and $t$ integral variables. The $t$-integral can be performed using the second Barnes' lemma ($\ref{secondlemma}$). Just like before the $t$-integral generates a $\Gamma(\epsilon)$, which cancels the one generated by the $\bar{z}$-integral above. Now that the divergences have canceled, we can smoothly take the $\epsilon \rightarrow 0$ limit. Applying this to the OPE correction case ($\ref{OPEcorrectionmaster}$), we will split the contribution into two parts (one coming from Mack polynomial term and one from Mack derivative term). The final expression involves the remaining integral in $s$,
\begin{align}
\begin{split} \label{mackterms}
\delta P^{(1)}_{\D,J}=&\alpha_{J}\sum_{m=0}^J \frac{(-1)^{m+1}}{(\l_1-m)_m^2} A_m(J,\D)\\
&\times \int_\mc ds\G(-s) \frac{(\l_1-m)_s(1-\bar{\l}_2)_s}{(1+\l_2-\bar{\l}_2)_{s+J-m}}  \frac{\Gamma(\frac{\beta-\tau}{2}-\l_1+m -1 -s)\Gamma(\frac{\tau}{2}+\l_1-m +1 +s)^2}{\Gamma(\frac{\beta+\tau}{2}+\l_1-m +1 +s)}\\
&\times [(H_{s+\l_1-m-1}- \pi \cot\pi(s+\l_1-m) -H_{s-\bar{\l}_2}+H_{-\l_2-J+m}+H_{-\bar{\l}_2}) ]\,.
\end{split}
\end{align}

The Mack derivative term is, 
\begin{align}\label{mackderv}
\begin{split}
\delta P^{(2)}_{\D,J}=&\frac{\alpha_J}{(d-\Delta-1)_J} \sum_{m=0}^J\sum_{n=0}^{J-m} \frac{(-1)^{m+n+1}}{(\l_1-m-n)_{m+n}^2} B_{m,n}(J,\D) \\
&\times \int_\mc ds\G(-s) \frac{(\l_1-m-n)_s(1-\bar{\l}_2)_s}{(1+\l_2-\bar{\l}_2)_{s+J-m-n}}  \frac{\Gamma(\frac{\beta-\tau}{2}-\l_1+m+n -1 -s)\Gamma(\frac{\tau}{2}+\l_1-m-n +1 +s)^2}{\Gamma(\frac{\beta+\tau}{2}+\l_1-m-n +1 +s)}\,.
\end{split}
\end{align}
In the above equations we have defined $\alpha_J$ as,
\begin{equation}
\alpha_J = 2\sin ^2\left[\left(\frac{\tau}{2}+\l_2\right) \pi\right]   \frac{\kappa_\beta}{C_0(\beta)} \frac{\G(2\l_1)\G(\beta)}{\G(\l_1)^2\G(\frac{\beta}{2})^2} \frac{1}{(d-2)_J}.     
\end{equation}
The $2\sin ^2\left[\left(\frac{\tau}{2}+\l_2\right) \pi\right]$ term comes from the double-discontinuity of the conformal block and its pre-factor: $\left(\frac{1-\bar{z}}{\bar{z}}\right)^\frac{\tau}{2} G^t_{\Delta,J}\vert_{\text{reg}}$. Now we will proceed with the s-integral.
 
		\subsection{Terms involving Mack polynomial}
		For simplicity of calculation we will split ($\ref{mackterms}$) into three parts. $I_1$ contains the integral with Harmonic numbers that depend on s,
		\begin{align}\label{I1integral}
			\begin{split}
				I_1=\alpha_J \sum_{m=0}^J (-1)^{m+1} \frac{A_m(J,\D)}{(\l_1-m)_m^2} \int_\mc &ds\G(-s) \frac{(\l_1-m)_s(1-\bar{\l}_2)_s}{(1+\l_2-\bar{\l}_2)_{s+J-m}}  \frac{\Gamma(\frac{\beta-\tau}{2}-\l_1+m -1 -s)}{\Gamma(\frac{\beta+\tau}{2}+\l_1-m +1 +s)}\\
				&\times \Gamma(\frac{\tau}{2}+\l_1-m +1 +s)^2\,[H_{s+\l_1-m-1} -H_{s-\bar{\l}_2}]\,.
			\end{split}
		\end{align}
		$I_2$ is an integral involving the Cotangent term,
		\begin{align}
			\begin{split}
				I_2=&\alpha_J\pi \sum_{m=0}^J \frac{(-1)^m}{(\l_1-m)_m^2} A_m(J,\D) \\
				&\int_\mc ds\G(-s) \frac{(\l_1-m)_s(1-\bar{\l}_2)_s}{(1+\l_2-\bar{\l}_2)_{s+J-m}}  \frac{\Gamma(\frac{\beta-\tau}{2}-\l_1+m -1 -s)\Gamma(\frac{\tau}{2}+\l_1-m +1 +s)^2}{\Gamma(\frac{\beta+\tau}{2}+\l_1-m +1 +s)}\\
				&\times \cot\pi(s+\l_1-m) \,.
			\end{split}
		\end{align}
		$I_3$ is the integral involving the remaining pieces,
		\begin{align}
			\begin{split}
				I_3=&\alpha_J \sum_{m=0}^J \frac{(-1)^{m+1}}{(\l_1-m)_m^2} A_m(J,\D)\,[H_{-\l_2-J+m} +H_{-\bar{\l}_2}]\\ &\times \int_\mc ds\G(-s) \frac{(\l_1-m)_s(1-\bar{\l}_2)_s}{(1+\l_2-\bar{\l}_2)_{s+J-m}}  \frac{\Gamma(\frac{\beta-\tau}{2}-\l_1+m -1 -s)\Gamma(\frac{\tau}{2}+\l_1-m +1 +s)^2}{\Gamma(\frac{\beta+\tau}{2}+\l_1-m +1 +s)}\,.
			\end{split}
		\end{align}
		We will perform these integrals separately and then add the contributions up together. Closing the contour to the right half plane gives rise to two series of poles in s,
		\begin{equation}
			s\in\mathbb{N} \qquad s\in\frac{\beta-\tau}{2}-\l_1+m -1 +\mathbb{N}
		\end{equation}
		These two infinite sum over residues again give rise to two ${}_4F_3$ hypergeometrics. 	
		The integral in $I_1$ involves Harmonic numbers therefore the sum over residues would involve Harmonic numbers in sum as well. To perform this sum we will employ the trick given in appendix (\ref{harmonicsumtrick}). The final result is,
		\begin{align}\label{I1spinj}
			\begin{split}
				I_1=&\alpha_J \sum_{m=0}^J \frac{(-1)^{m+1}}{(\l_1-m)_m^2} A_m(J,\D) \frac{\G(1+\l_2-\bar{\l}_2)}{\G(\l_1-m) \G(1-\bar{\l}_2)}\\
				\Bigg[& C_m(J,\Delta) \Bigg( \,\,\, _4F_3\bigg[\begin{matrix}1-h+\l_1,\l_1-m,\frac{\tau}{2}+\l_1-m+1,\frac{\tau}{2}+\l_1-m+1\\ -h+2\l_1-m+1,\frac{-\beta+\tau}{2}+\l_1-m+2,\frac{\beta+\tau}{2}+\l_1-m+1\end{matrix};1\bigg]   \\ &( H_{\frac{J+\Delta}{2}-m-1} - H_{-h+\frac{J+\Delta}{2}} ) -\mathcal{G}_1 \,\Bigg) \\
				&+D_m(J,\Delta)\Bigg(\,\,\,_4F_3\bigg[\begin{matrix}\frac{\beta }{2},\frac{\beta }{2},\frac{\beta -\tau}{2}-1,\frac{\beta -\tau}{2}-h+m\\ \beta,\frac{\beta-\tau }{2}-\l_1+m,-h+\frac{\beta  - \tau }{2}+\l_1\end{matrix};1\bigg] (H_{\frac{\beta -\tau}{2}-2} - H_{\frac{\beta -\tau}{2}-h+m-1})+\mathcal{G}_2\,\Bigg)\Bigg],
			\end{split}
		\end{align}
		where we have defined,
		\begin{align}
			\begin{split}
				C_m(J,\Delta) = &\frac{\Gamma \left(\l_1 - h+1 \right) \Gamma \left(\l_1 - m\right)
					\Gamma \left(\frac{\tau}{2}+\l_1+1 - m\right)^2 \Gamma \left(m-1+\frac{\beta -\tau }{2}-\l_1 \right)}{\Gamma
					(-h-m+2\l_1+1) \Gamma \left(\frac{\beta +\tau }{2}+\l_1 -m+1\right)},  \\
				D_m(J,\Delta) = &\frac{\Gamma \left(\frac{\beta }{2}\right)^2 \Gamma \left(\frac{\beta -\tau }{2} -1 \right) \Gamma
					\left(- h+ m+\frac{\beta -\tau }{2} )\right) \Gamma \left(\frac{\tau-\beta 
					}{2}+\l_1 -m+1\right)}{\Gamma (\beta ) \Gamma \left(\frac{\beta  -\tau }{2}+\l_1 -h\right)} \, ,
			\end{split}
		\end{align}
		to make these expressions more compact. The functions $\mathcal{G}_1$ and $\mathcal{G}_2$ are the  Kamp\'e de F\'eriet-like functions defined in (\ref{kampeG}).

For the remaining terms, performing the integral is straightforward and we obtain, 
\begin{align}
\begin{split}
I_2=&\pi \alpha_J \sum_{m=0}^J \frac{(-1)^m}{(\l_1-m)_m^2} A_m(J,\D) \frac{\G(1+\l_2-\bar{\l}_2)}{\G(\l_1-m) \G(1-\bar{\l}_2)}\Bigg( C_m(J,\Delta) \,\, \cot\left[\pi\left(\l_1-m\right)\right]\,\,\, \\
&\times _4F_3\bigg[\begin{matrix}1-h+\l_1,\l_1-m,\frac{\tau}{2}+\l_1-m+1,\frac{\tau}{2}+\l_1-m+1\\ -h+2\l_1-m+1,\frac{-\beta+\tau}{2}+\l_1-m+2,\frac{\beta+\tau}{2}+\l_1-m+1\end{matrix};1\bigg] \\
&+D_m(J,\Delta) \,\, \cot\left[\pi\left(\frac{\beta-\tau}{2}\right)\right]\,\,\,_4F_3\bigg[\begin{matrix}\frac{\beta }{2},\frac{\beta }{2},\frac{\beta -\tau}{2}-1,\frac{\beta -\tau}{2}-h+m\\ \beta,\frac{\beta-\tau }{2}-\l_1+m,-h+\frac{\beta  - \tau }{2}+\l_1\end{matrix};1\bigg]  \Bigg)\, ,
\end{split}
\end{align}
and,
\begin{align}
\begin{split}
I_3=&\alpha_J \sum_{m=0}^J \frac{(-1)^{m+1}}{(\l_1-m)_m^2} A_m(J,\D) \frac{\G(1+\l_2-\bar{\l}_2)}{\G(\l_1-m) \G(1-\bar{\l}_2)}\,[H_{-\l_2-J+m} +H_{-\bar{\l}_2}]\\
\Bigg[& C_m(J,\Delta) \,\, \,\,\, _4F_3\bigg[\begin{matrix}1-h+\l_1,\l_1-m,\frac{\tau}{2}+\l_1-m+1,\frac{\tau}{2}+\l_1-m+1\\ -h+2\l_1-m+1,\frac{-\beta+\tau}{2}+\l_1-m+2,\frac{\beta+\tau}{2}+\l_1-m+1\end{matrix};1\bigg]  \\
&+D_m(J,\Delta) \,\, \,\,\,_4F_3\bigg[\begin{matrix}\frac{\beta }{2},\frac{\beta }{2},\frac{\beta -\tau}{2}-1,\frac{\beta -\tau}{2}-h+m\\ \beta,\frac{\beta-\tau }{2}-\l_1+m,-h+\frac{\beta  - \tau }{2}+\l_1\end{matrix};1\bigg]  \Bigg].
\end{split}
\end{align}
		
\subsection{Terms involving derivative of Mack polynomial}
Performing the s-integral in (\ref{mackderv}) results in the following piece of the correction,
		\begin{align}
			\begin{split}
				\delta P&^{(2)}_{\D,J}=\frac{\alpha_J}{(d-\Delta-1)_J} \sum_{m+n\leq J} \frac{(-1)^{m+n+1}}{(\l_1-m-n)_{m+n}^2} B_{m,n}(J,\D) \frac{\G(1+\l_2-\bar{\l}_2)}{\G(\l_1-m-n) \G(1-\bar{\l}_2)}\, \\
				&\times \Bigg( C_{m+n}(J,\Delta) \,\, \,\, _4F_3\bigg[\begin{matrix}1-h+\l_1,\l_1-m,\frac{\tau}{2}+\l_1-m+1,\frac{\tau}{2}+\l_1-m+1\\ -h+2\l_1-m+1,\frac{-\beta+\tau}{2}+\l_1-m+2,\frac{\beta+\tau}{2}+\l_1-m+1\end{matrix};1\bigg]  \\
				&+D_{m+n}(J,\Delta) \,\, \,\,\,_4F_3\bigg[\begin{matrix}\frac{\beta }{2},\frac{\beta }{2},\frac{\beta -\tau}{2}-1,\frac{\beta -\tau}{2}-h+m\\ \beta,\frac{\beta-\tau }{2}-\l_1+m,-h+\frac{\beta  - \tau }{2}+\l_1\end{matrix};1\bigg]  \Bigg). 
			\end{split} 
		\end{align}
		This term has a double sum in $m$ and $n$ variables with the constraint that $m+n<J$.
\subsection{Total correction to OPE coefficients}
		The net contribution to the OPE correction for a general spin-$J$ exchange is the sum of all the above terms,
		\begin{align}
		\nonumber
			\delta P_{\D,J}&= \delta P^{(1)}_{\D,J}+\delta P^{(2)}_{\D,J} \\
			&=I_1 +I_2+I_3+\delta P^{(2)}_{\D,J}    
		\end{align}
		We can recover the scalar exchange OPE correction by setting $m=J=0$ in the above equation. From ($\ref{I1spinj}$) we obtain,
		\begin{align}
			\begin{split}
				&I_1=-\alpha_0\frac{\G(1-h+\Delta)}{\G(1-h+\frac{\Delta}{2}) \G(\frac{\Delta}{2})}\Bigg( C_0(0,\Delta)\Bigg[ \pFq{4}{3}{\frac{\Delta}{2},1-h+\frac{\Delta}{2},1+\frac{\Delta+\tau}{2},1+\frac{\Delta+\tau}{2}}{1-h+\Delta,2-\frac{\beta-\Delta-\tau}{2},1+\frac{\beta+\Delta+\tau}{2}}{1}\\&\times( H_{\frac{\Delta}{2}-1} - H_{-h+\frac{\Delta}{2}}) -\mathcal{G}_1 \Big] +D_0(0,\Delta)\\& 
				\times \Big[ \pFq{4}{3}{\frac{\beta}{2},\frac{\beta}{2},-1+\frac{\beta+\tau}{2},-h+\frac{\beta+\tau}{2}}{\beta,\frac{\beta-\tau-\Delta}{2},-h+\frac{\beta+\Delta-\tau}{2}}{1}(H_{\frac{\beta -\tau}{2}-2} - H_{\frac{\beta -\tau}{2}-h-1})+ \mathcal{G}_2\Big]\Bigg)
			\end{split}
		\end{align}
		In the above equation $\mathcal{G}_1$ and $\mathcal{G}_2$ are (\ref{kampeG}) with $m=J=0$.
		The result for the cotangent contribution is,
		\small{
			\begin{align}\label{i2scalar}
				\begin{split}
					I_2=&\pi\alpha_0\frac{\G(1-h+\Delta)}{\G(1-h+\frac{\Delta}{2}) \G(\frac{\Delta}{2})}\Bigg(D_0(0,\Delta) \cot \bigg(\frac{\pi(\beta-\tau)}{2}\bigg) \ \pFq{4}{3}{\frac{\beta}{2},\frac{\beta}{2},-1+\frac{\beta+\tau}{2},-h+\frac{\beta+\tau}{2}}{\beta,\frac{\beta-\tau-\Delta}{2},-h+\frac{\beta+\Delta-\tau}{2}}{1} \\&+ C_0(0,\Delta)\cot\bigg(\frac{\pi\Delta}{2}\bigg)\ \pFq{4}{3}{\frac{\Delta}{2},1-h+\frac{\Delta}{2},1+\frac{\Delta+\tau}{2},1+\frac{\Delta+\tau}{2}}{1-h+\Delta,2-\frac{\beta-\Delta-\tau}{2},1+\frac{\beta+\Delta+\tau}{2}}{1}\Bigg) \,.
				\end{split}
			\end{align}
		}
		The constant terms are,
		\small{
			\begin{align}\label{i3scalar}
				\begin{split}
					I_3=&-\alpha_0\frac{\G(1-h+\Delta)}{\G(1-h+\frac{\Delta}{2}) \G(\frac{\Delta}{2})}\Bigg(D_0(0,\Delta)  \,\, \pFq{4}{3}{\frac{\beta}{2},\frac{\beta}{2},-1+\frac{\beta+\tau}{2},-h+\frac{\beta+\tau}{2}}{\beta,\frac{\beta-\tau-\Delta}{2},-h+\frac{\beta+\Delta-\tau}{2}}{1} \\&+ C_0(0,\Delta) \pFq{4}{3}{\frac{\Delta}{2},1-h+\frac{\Delta}{2},1+\frac{\Delta+\tau}{2},1+\frac{\Delta+\tau}{2}}{1-h+\Delta,2-\frac{\beta-\Delta-\tau}{2},1+\frac{\beta+\Delta+\tau}{2}}{1}\Bigg) \\
					&  \times (H_{-\frac{\D}{2}}+H_{\frac{\D}{2}-h})\,.
				\end{split}
			\end{align}
		}
	Putting all the pieces together we obtain the total correction to the OPE coefficient to be,
		\begin{equation}
			\delta P^t_{\D,0} =I_1+I_2+I_3.
		\end{equation}
There is no Mack derivative term $\delta P^{(2)}_{\D,0}$ for the scalar exchange case. Thus for the scalar the correction to the OPE coefficient essentially comes from the finite part (excluding the $\log$ term) of the measure. For spin case, the additional contribution comes from $\d P^{(2)}_{\D,J}$ part. 
\subsection{Special cases}
The expression for OPE coefficients undergo simplifications in even dimensions. The terms involving Kamp\'e de F\'eriet-like double sums reduce to ${}_4F_3$ Hypergeometrics. Since those terms are generated by the $I_1$ integral, their reduction can be easily seen from the integral representation of $I_1$,
		\begin{align}
			\begin{split}
				I_1=\alpha_J \sum_{m=0}^J \frac{(-1)^{m+1}}{(\l_1-m)_m^2} \int_\mc &ds\G(-s) \frac{(\l_1-m)_s(1-\bar{\l}_2)_s}{(1+\l_2-\bar{\l}_2)_{s+J-m}}  \frac{\Gamma(\frac{\beta-\Delta-J-\tau}{2}+m -1 -s)\Gamma(\frac{\Delta+J+\tau}{2}-m +1 +s)^2}{\Gamma(\frac{\beta+\Delta+J+\tau}{2}-m +1 +s)}\\
				&\times A_m(J,\D)\,[H_{s+\l_1-m-1} -H_{s-\bar{\l}_2}]\,.
			\end{split}
		\end{align}
		The double sums were generated by the derivatives of ${}_4F_3$ hypergeometric with respect to their parameters. In even dimensions the Harmonic number parameters are integer separated,
		\begin{equation}\label{harmdiff}
			H_{s+\l_1-m-1} -H_{s-\bar{\l}_2} \rightarrow H_{s+\frac{\Delta + J}{2}-m-1} -H_{s+\frac{\Delta + J}{2}-h} \, ,
		\end{equation}
		In the above equation $m$ is an integer while $h$ is an integer in even dimensions. In this case, the difference of Harmonic numbers reduces to a simple functions of $s$ using the Harmonic number recursion relations. For simplicity we will consider the case of scalars where ($m=J=0$). The difference (\ref{harmdiff}) vanishes in two dimensions and in four dimensions it simplifies to,
		\begin{equation}
			H_{s+\frac{\Delta}{2}-1} -H_{s+\frac{\Delta}{2}-2}=\frac{1}{s+\frac{\Delta}{2}-1} \, .
		\end{equation}
\subsubsection{Two dimensions}
Since $h$=1 the integral for $I_1$ vanishes in two dimensions,
\begin{equation}
			I_1^{2d}=0
\end{equation}
 For the other integrals ($I_2$ and $I_3$) we can directly take the final result for the scalar case from ($\ref{i2scalar}$) and ($\ref{i3scalar}$) and substitute the values of $h$. Let us first define the following functions:
 \begin{equation}
 f_1(h)=\frac{\G(1-h+\Delta)}{\G(1-h+\frac{\Delta}{2}) \G(\frac{\Delta}{2})}\,,\ C_0(0,\Delta)=\frac{\Gamma \left(\frac{\Delta +\tau +2}{2}\right)^2 \Gamma \left(\frac{\beta -\Delta -\tau -2}{2} \right)}{\Gamma \left(\frac{\beta +\Delta +\tau +2}{2}\right)} \,,
 \end{equation}
 \begin{align}
 \begin{split}
 f_2(h)=&D_0(0,\Delta)\frac{\G(1-h+\Delta)}{\G(1-h+\frac{\Delta}{2}) \G(\frac{\Delta}{2})}\\&=\frac{\Gamma \left(\frac{\beta }{2}\right)^2 \Gamma (-h+\Delta +1) \Gamma \left(\frac{-\beta +\Delta +\tau +2}{2}\right) \Gamma \left(\frac{+\beta -\tau -2}{2}\right) \Gamma \left(\frac{-2 h+\beta -\tau }{2}\right)}{\Gamma (\beta ) \Gamma \left(\frac{\Delta
 	}{2}\right) \Gamma \left(\frac{-2 h+\Delta +2}{2} \right) \Gamma \left(\frac{-2 h+\beta +\Delta -\tau }{2} \right)}\, .
 \end{split}
 \end{align}
 The correction to OPE coefficient for a scalar in two dimension becomes,
 		\small{
 	\begin{align}
 	\begin{split}
 	\delta P_{\Delta,0}^{2d}=&\alpha_0\Big( f_2(1)\pFq{4}{3}{\frac{\beta}{2},\frac{\beta}{2},-1+\frac{\beta+\tau}{2},-1+\frac{\beta+\tau}{2}}{\beta,\frac{\beta-\tau-\Delta}{2},-1+\frac{\beta+\Delta-\tau}{2}}{1}(\pi \cot(\pi \Delta/2)-H_{-\frac{\D}{2}}-H_{\frac{\D}{2}-1})\\
 	&+f_1(1)\pFq{4}{3}{\frac{\Delta}{2},\frac{\Delta}{2},1+\frac{\Delta+\tau}{2},1+\frac{\Delta+\tau}{2}}{\Delta,2-\frac{\beta-\Delta-\tau}{2},1+\frac{\beta+\Delta+\tau}{2}}{1}(\pi \cot(\pi (\beta-\tau)/2)-H_{-\frac{\D}{2}}-H_{\frac{\D}{2}-1})\Big) \,,
 	\end{split}
 	\end{align}
 }
\subsubsection{Four dimensions}
For four dimensions, the $I_1$ integral simplifies to,
\begin{align}
	\begin{split}
				I_1^{4d}=-\alpha_0 \int_\mc &ds\G(-s) \frac{(\l_1)_s(1-\bar{\l}_2)_s}{(1+\l_2-\bar{\l}_2)_{s}}  \frac{\Gamma(\frac{\beta-\Delta-\tau}{2} -1 -s)\Gamma(\frac{\Delta+\tau}{2} +1 +s)^2}{\Gamma(\frac{\beta+\Delta+\tau}{2} +1 +s)}
				\,\frac{1}{s+\frac{\Delta}{2}-1}\,.
			\end{split}
\end{align}
		This integral can be easily evaluated and results in two ${}_4F_3$ hypergeometric functions only and no Kamp\'e de F\'eriet-like double sums. 
		
		\begin{align}
			\begin{split}
				\delta P_{\Delta,0}^{4d}=&\alpha_0\Big(f_1(2)\pFq{4}{3}{\frac{\Delta}{2},-1+\frac{\Delta}{2},1+\frac{\Delta+\tau}{2},1+\frac{\Delta+\tau}{2}}{-1+\Delta,2-\frac{\beta-\Delta-\tau}{2},1+\frac{\beta+\Delta+\tau}{2}}{1}(\pi \cot(\pi \Delta/2)-H_{-\frac{\D}{2}}-H_{\frac{\D}{2}-1}) \\
				&-f_1(2)\pFq{4}{3}{-1+\frac{\Delta}{2},-1+\frac{\Delta}{2},1+\frac{\Delta+\tau}{2},1+\frac{\Delta+\tau}{2}}{-1+\Delta,2-\frac{\beta-\Delta-\tau}{2},1+\frac{\beta+\Delta+\tau}{2}}{1}(\frac{2}{\Delta-2}) \\
				&+f_2(2)\pFq{4}{3}{\frac{\beta}{2},\frac{\beta}{2},-1+\frac{\beta+\tau}{2},-2+\frac{\beta+\tau}{2}}{\beta,\frac{\beta-\tau-\Delta}{2},-2+\frac{\beta+\Delta-\tau}{2}}{1}(\pi \cot(\pi (\beta-\tau)/2)-H_{-\frac{\D}{2}}-H_{\frac{\D}{2}-1}) \\
				&-f_2(2)\pFq{4}{3}{\frac{\beta}{2},\frac{\beta}{2},-2+\frac{\beta+\tau}{2},-2+\frac{\beta+\tau}{2}}{\beta,\frac{\beta-\tau-\Delta}{2},-2+\frac{\beta+\Delta-\tau}{2}}{1}(\frac{2}{\beta-\tau-4})\Big)\, .
			\end{split}
		\end{align}
		The extra ${}_4F_3$ hypergeometrics in the four dimensions case compared to two dimensions are generated from the $I_1$ integral (which vanished in 2d). 

\section{Conclusions}
In this paper we have considered the single block contributions to the anomalous dimensions and OPE coefficients of operators $[\mo_1\mo_2]_{\D,J}$ with dimensions $\D=\D_1+\D_2+J$. In order to achieve that, we have used the integral (Mellin) representation of the blocks which make the analysis and results fairly general. On the one hand it is democratic with respect to space time dimensions, thus making the computation uniform both in even and odd dimensions, specially where the closed form of the conformal blocks is not available. Further, the Mellin integral makes the additional contributions (coming from the lower limit of the $z$ integral) explicit, making it possible to re-sum correctly for any finite $\b$. 

We would like to highlight some salient points from this work and  future directions.
\begin{itemize} 
\item The anomalous dimensions and the corrections to the OPE coefficients can be written as an exact function of the conformal spin ($\b$) in terms of Wilson polynomials for each exchange contribution in the $t-$channel. These Wilson polynomials are the generalizations of the residues of the $6j-$symbols recently discussed in \cite{Liu:2018jhs}. 
\item The computations of the anomalous dimensions and corrections to the OPE coefficients follow uniformly from the integral representation of the physical blocks by extracting the $\log$ and regular terms respectively.
\item The closed form expressions create a possibility for numerical exploration along with a proper handling of the associated error estimates.
\item The computation for operators $[\mo_1(\pd^2)^n\mo_2]_{\D,J}$ with $\D=\D_1+\D_2+J+2n$ can be handled in the exact same fashion as for the $n=0$ case. In this case, one needs to consider the descendant contributions from the Mellin representation of the blocks we neglected in this work\footnote{We focussed on the $t=0$ poles. For the descendants, we need to take into account $t=n$ poles in \eqref{intr}}. Along with the descendant contributions from the block, one also needs to expand the kernel in the inversion formula \eqref{exactC}. For our case, we considered,
\be
\lim_{z\rightarrow0} G_{J+d-1,\D+1-d}(z,\bz)\sim z^{\frac{J-\D}{2}}k_\b(\bz)+\dots\,,
\ee
where $\dots$ terms become revelant in the subleading orders ({\it i.e.} for $n>0$ cases). As a result, there is a mixing problem involved at the subleading orders. For example $z^i$ term coming from the kernel and $z^j$ term can combine to $z^n$ where $n=i+j$. This can be interpreted in the following way. At any subleading order, we have contributions from the primary $[\mo_1(\pd^2)^n\mo_2]_{\D,J}$ and the $m-$th descendant of the primary $[\mo_1(\pd^2)^{n-m}\mo_2]_{\D,J}$. It would be interesting to see how these contributions can be disentangled\footnote{We thank Aninda Sinha for discussion on this point.}.
\item The obvious generalization of our approach is to consider the extension to external operators with spin. Using the recent developments in \cite{Sleight:2018epi}, this can be achieved. Then we might be able to obtain the anomalous dimensions and corrections to OPE coefficients of more generic ``double-twist" operators of the form $[\mo_{J_1}\mo_{J_2}]_{\D,J}$. 
\item Finally it would be nice to understand the full consequence of the inversion formula along the lines of section 4.3.1 and 4.3.2 of \cite{Caron-Huot:2017vep}. Once we know the contributions of the individual blocks not only in the $z\rightarrow0$ limit but for all $z,\bz$, we can analyze the constraints from finite subtractions. We would like to add that an open arena is still to understand the implications of the inversion formula for deformations of CFTs like considered in \cite{Sen:2017gfr} and for the case of the bulk duals at finite coupling considered in \cite{Carmi:2018qzm}. 
\end{itemize}

\section{Acknowledgements}
KS would like to thank Simon Caron-Huot for discussions and University of McGill for hospitality, where the final stages of the work was completed. SG would like to thank his advisors Katrin and Melanie Becker for partially supporting him. SKK would like to thank his advisor Christopher Pope for supporting him through DOE grant DE-FG02-13ER42020. KS is partially supported by World Premier Research Center Initiative (WPI) initiative, MEXT Japan at Kavli IPMU, World Premier Research Center Initiative (WPI) the University of Tokyo. CC is supported in part by the Danish National Research Foundation (DNRF91).

\appendix

\section{Integral representation}\label{intrep}

We will start with the integral representation of the conformal blocks following \cite{Mack:2009mi, Dolan:2000ut,Dolan:2011dv}. The integral representation for the four point function $\la\mo_1\mo_2\mo_3\mo_4\ra$ in the OPE decomposition $\mo_1\times\mo_2$ and $\mo_3\times\mo_4$ due to the exchange of an operator $\mo_{J,\D}$, is given by,
\begin{align}\label{intr}
\begin{split}
f_{\D,J}(u,v)=\frac{1}{\g_{\l_1,a}\g_{\bar{\l_1},b}}\int_{\mathcal{C}} ds dt\ \G(\l_2-s)&\G(\bar{\l}_2-s)\G(-t)\G(-t-a-b)\\
&\times\G(s+t+a)\G(s+t+b)\mathcal{P}_{\D,J}(s,t,a,b)u^s v^t\,,
\end{split}
\end{align}
where we have stripped off the overall kinematical factors. The contour $\mathcal{C}$ extends from $\g-i\infty$ to $\g+i\infty$ where following \cite{Dolan:2000ut, Dolan:2011dv}, 
\be
\text{Re} (s)<\l_2,\bar{\l}_2\,, \ \text{Re} (t)<0,-a-b\,,\ \text{Re} (c)<a,b\,,
\ee
and for the $t$ integral, $-a-s,-b-s<\g<0,-a-b$. In general,
\be
f_{\D,J}(u,v)=\frac{1}{k_{d-\D,J}}\frac{\g_{\l_1,a}}{\g_{\bar{\l}_1,b}}G_{\D,J}(u,v)+\frac{1}{k_{\D,J}}\frac{\g_{\bar{\l}_1,a}}{\g_{\l_1,a}}G_{d-\D,J}(u,v)\,,
\ee
is a linear combination of the physical block and the shadow respectively from the $s=\l_2+n$ and $s=\bar{\l}_2+n$ poles. We have also used the definitions,
\begin{align}
\l_1=\frac{\D+J}{2}\,, \ \bar{\l}_1=\frac{d-\D+J}{2}\,,\\
\l_2=\frac{\D-J}{2}\,, \ \bar{\l}_2=\frac{d-\D-J}{2}\,.
\end{align}
$a=\frac{\D_{21}}{2}$ and $b=\frac{\D_{34}}{2}$ where $\D_{ij}=\D_i-\D_j$,  
and,
\be
k_{\D,J}=\frac{1}{(\D-1)_J}\frac{\G(d-\D+J)}{\G(\D-h)}\,,\ \ \g_{x,y}=\G(x+y)\G(x-y)\,, \ \text{and}\ h=d/2\,.
\ee
$d$ is the space-time dimension. $\mathcal{P}_{\D,J}(s,t,a,b)$ is the Mack polynomial given by,
\begin{align}\label{mackp}
\begin{split}
\mathcal{P}_{\D,J}(s,t,a,b)=&\frac{1}{(d-2)_J}\sum_{m+n+p+q=J}\frac{J!}{m!n!p!q!}(-1)^{p+n}(2\bar{\l}_2+J-1)_{J-q}(2\l_2+J-1)_n(\bar{\l}_1+a-q)_q\\
&\times(\bar{\l}_1+b-q)_q(\l_1+a-m)_m(\l_1+b-m)_m(d-2+J+n-q)_q(h-1)_{J-q}\\
&\times(h-1+n+a+b)_p(\l_2-s)_{p+q}(-t)_n\,.
\end{split}
\end{align}
The Mack polynomial and it's derivative for $a=b=0$ are,
\begin{align}\label{macksimp}
\begin{split}
\mathcal{P}_{\D,J}(s,0)=\frac{(d-\D-1)_J}{(d-2)_J}\sum_{m=0}^J A_m(J,\D)(\l_2-s)_{J-m}\,, \\
\mathcal{P}_{\D,J}'(s,0)=\frac{(d-\D-1)_J}{(d-2)_J}\sum_{1\leq m+n\leq J} B_{m,n}(J,\D)(\l_2-s)_{J-m-n}\,,
\end{split}
\end{align}
with 
\begin{align}\label{coeffab}
\begin{split}
A_m(J,\D)=&\frac{(1+J-m)_m(h-1)_m(\l_1-m)_m^2(\bar{\l}_2+m)_{J-m}^2(2h-2+m)_{J-m}(2\bar{\l}_2-1+J)_m}{\G(m+1)(d-\D-1)_J}\\
&\times {}_4F_3\bigg[\begin{matrix}h-1,h+m-1,m-J,2\bar{\l}_2+J-1+m\\ 2h-2+m,\bar{\l}_1-J+m,\bar{\l}_1-J+m\end{matrix};1\bigg]\,,\\
B_{m,n}(J,\Delta) =& \frac{(-1)^{n+1} (1-\delta_{0,n})(\Delta-1)_{n}}{n}  (\l_1 -m)^2_m(1+J-m-n)_{m+n} \,\\&\times\frac{(h-1)_{m+n}(h+m+n-\l_1)_{J-m-n}^2 \Gamma (2 h+m+2n-2)_{J-m-n}(2\bar{\l}_2-1+J)_{m+n}}{\G(m+1)(d-\D-1)_J}\\
&\times {}_4F_3\bigg[\begin{matrix}-1+h+n,-1+h+m+n,-J+m+n,-1+2h+m+n-\D\\ -2+2h+m+2n,h+m+n-\l_1,h+m+n-\l_1\end{matrix};1\bigg]\, .
\end{split} 
\end{align}
In what follows, we will select a scheme to write down the integral representation corresponding to the physical block itself in $d$ dimensions. 
 We consider the integral representation of the conformal blocks for a general spin exchange, given by,
\be\label{intr1}
G_{\D,J}(1-z,1-\bz)=\frac{k_{d-\D,J}}{\g_{\l_1}^2}\int_\mc ds dt\ \G(\l_2-s)\G(\bar{\l}_2-s)\G(-t)^2\G(s+t)^2 P_{\D,J}(s,t) (z\bz)^t [(1-z)(1-\bz)]^s\,.
\ee
By closing the contour on the {\it rhs} of the complex $s-$plane, one finds there are two sets of poles at $s=\l_2+n$ and $s=\bar{\l}_2+n$ characterizing the physical and the shadow blocks respectively. The idea is to remove the contribution of the shadow block completely. This is achieved by multiplying the integral representation by a phase,
\be
p(s)=\frac{\sin \pi(\bar{\l}_2-s)}{\sin \pi(\l_2-\bar{\l}_2)}\frac{\sin \pi s}{\sin \pi\l_1}\,,
\ee
such that the shadow poles are now completely removed. The phase satisfies the shift symmetry property such that for $s\rightarrow s\pm k$, $p(s\pm k)=p(s)$. We then write the modified integral definition for the physical block as,
\begin{align}
\begin{split}
G_{\D,J}(1-z,1-\bz)=\frac{k_{d-\D,J}}{\g_{\l_1}^2}\int_\mc ds dt\ &\G(\l_2-s)\G(\bar{\l}_2-s)\G(-t)^2\G(s+t)^2 p(s)\\
&\times P_{\D,J}(s,t) (z\bz)^t [(1-z)(1-\bz)]^s\,.
\end{split}
\end{align}
We will be mainly interested in the $z\rightarrow0$ limit of the block for the leading corrections to the dimension and the OPE coefficients discussed in the paper. Notice that for this limit, only the contribution from the $t=0$ pole suffices. Moreover, we can rewrite $\bar{z}^{s+t}$ as,
\begin{equation}
\bz^{s+t}=\sum_{k=0}^\infty \frac{(-1)^k}{k!}(s+t)_k \bigg(\frac{1-\bz}{\bz}\bigg)^k\,,
\end{equation}
Thus, after taking the $t=0$ pole,  
\begin{align}
\begin{split}
\lim_{z\rightarrow0}G_{\D,J}(1-z,1-\bz)=&\frac{k_{d-\D,J}}{\g_{\l_1}^2}\int_\mc ds\ \G(\l_2-s)\G(\bar{\l}_2-s)\G(s)p(s)\\
&\times\sum_{k=0}^\infty\frac{(-1)^k}{k!}\G(s+k) \bigg(\frac{1-\bz}{\bz}\bigg)^{s+k}\bigg[ P_{\D,J}(s,0)(\log z+H_{s-k-1}+H_{s-1})\\
& + \mathcal{P}_{\D,J}'(s,0)]\,,
\end{split}
\end{align}
where $H_n$ is the Harmonic number $H(n)$. Before separating out the contributions to the anomalous dimensions and the OPE coefficients, we will perform a succession of shifts in the $s-$variable (mainly for the convenience of the computations to be followed). First we shift $s\rightarrow s-k$, such that, 
\begin{align}
\begin{split}
\lim_{z\rightarrow0}G_{\D,J}(1-z,1-\bz)=&\frac{k_{d-\D,J}}{\g_{\l_1}^2}\int_\mc ds\ \G(s)p(s)\bigg(\frac{1-\bz}{\bz}\bigg)^{s}\\
&\times\sum_{k=0}^\infty\frac{(-1)^k}{k!}\G(\l_2-s+k)\G(\bar{\l}_2-s+k)\G(s-k) \bigg[P_{\D,J}(s-k,0)\\
&\times(\log z+H_{s-k-1}+H_{s-1}) + \mathcal{P}_{\D,J}'(s-k,0)]\,.
\end{split}
\end{align}
The forms of the Mack polynomial and its derivative is given in \eqref{macksimp} along with the coefficients in \eqref{coeffab}. Plugging in those simplifications, we find, 
\begin{align}
\begin{split}
\lim_{z\rightarrow0}G_{\D,J}(1-z,1-\bz)=&\frac{\G(\D+J)}{\G(\frac{\D+J}{2})^4\G(h-\D)(d-2)_J}\int_\mc ds\ \G(s)p(s)\bigg(\frac{1-\bz}{\bz}\bigg)^{s}\\
&\times\sum_{k=0}^\infty\frac{(-1)^k}{k!}\G(\l_2-s+k)\G(\bar{\l}_2-s+k)\G(s-k) \bigg[\sum_{m=0}^J A_m(J,\D)(\l_2-s+k)_{J-m}\\
&\times(\log z+H_{s-k-1}+H_{s-1}) +\sum_{1\leq m+n\leq J} B_{m,n}(J,\D)(\l_2-s+k)_{J-m-n}\bigg]\,.
\end{split}
\end{align}
Based on the above separation, we can identify the coefficients of the $\log$ and regular terms as,
\begin{align}\label{logreg}
\begin{split}
\lim_{z\rightarrow0}G_{\D,J}(1-z,1-\bz)\bigg|_{\log z}=&\frac{\G(\D+J)}{\G(\frac{\D+J}{2})^4\G(h-\D)(d-2)_J} \sum_{m=0}^J A_m(J,\D)\int_\mc ds\ \G(s)p(s)\bigg(\frac{1-\bz}{\bz}\bigg)^{s}\\
&\times\sum_{k=0}^\infty\frac{(-1)^k}{k!}\G(\l_2-s+k)\G(\bar{\l}_2-s+k)\G(s-k)(\l_2-s+k)_{J-m}\,,\\
\lim_{z\rightarrow0}G_{\D,J}(1-z,1-\bz)\bigg|_{\text{ reg}}=&\frac{\G(\D+J)}{\G(\frac{\D+J}{2})^4\G(h-\D)(d-2)_J}\int_\mc ds\ \G(s)p(s)\bigg(\frac{1-\bz}{\bz}\bigg)^{s}\\
&\times\sum_{k=0}^\infty\frac{(-1)^k}{k!}\G(\l_2-s+k)\G(\bar{\l}_2-s+k)\G(s-k) \bigg[\sum_{m=0}^J A_m(J,\D)(\l_2-s+k)_{J-m}\\
&\times(H_{s-k-1}+H_{s-1}) +\sum_{1\leq m+n\leq J} B_{m,n}(J,\D)(\l_2-s+k)_{J-m-n}\bigg]\,.
\end{split}
\end{align}
This separation will form the starting point of discussion in the main text. However, as it stands \eqref{logreg} is still not ready in its final form to proceed with calculations. To put this in its final form, we will have to perform the $k-$sum now. As it stands, 
\begin{align}
\begin{split}
&\sum_{k=0}^\infty\frac{(-1)^k}{k!}\G(\l_2-s+k)\G(\bar{\l}_2-s+k)\G(s-k)(\l_2-s+k)_{J-m}\\
&=\frac{\pi}{\sin\pi s}\frac{\G(\l_1-m-s)\G(1-J+m+s-\l_2-\bar{\l}_2)\G(\bar{\l}_2-s)}{\G(1-\l_1+m)\G(1-\bar{\l}_2)}\,,
\end{split}
\end{align}
within the chosen domain of the $s-$contour. Similarly, 
\begin{align}
\begin{split}
&\sum_{k=0}^\infty\frac{(-1)^k}{k!}\G(\l_2-s+k)\G(\bar{\l}_2-s+k)\G(s-k)(\l_2-s+k)_{J-m}(H_{s-k-1}+H_{s-1})\\
=&\frac{\pi}{\sin\pi s}\frac{\G(\l_1-m-s)\G(1-J+m+s-\l_2-\bar{\l}_2)\G(\bar{\l}_2-s)}{\G(1-\l_1+m)\G(1-\bar{\l}_2)}(H_{s-1}-\pi\cot\pi s+H_{m-\l_1}\\
&-H_{s-\l_1+m-\bar{\l}_2}+H_{-\bar{\l}_2})\,,
\end{split}
\end{align}
We will now proceed with each of these terms separately. 

\subsection{$\log$ term}
Consider the integral representation of the $\log z$ term. After the $k-$summation, we get, 
\begin{align}
\begin{split}
\lim_{z\rightarrow0}G_{\D,J}(1-z,1-\bz)\bigg|_{\log z}=&\frac{\G(\D+J)}{\G(\frac{\D+J}{2})^4\G(h-\D)(d-2)_J} \sum_{m=0}^J A_m(J,\D)\int_\mc ds\ \G(s)p(s)\bigg(\frac{1-\bz}{\bz}\bigg)^{s}\\
&\times\frac{\pi}{\sin\pi s}\frac{\G(\l_1-m-s)\G(1-J+m+s-\l_2-\bar{\l}_2)\G(\bar{\l}_2-s)}{\G(1-\l_1+m)\G(1-\bar{\l}_2)}\,,
\end{split}
\end{align}
The choice of the phase factor now becomes more transparent. We can write, 
\be
p(s)\frac{\pi}{\sin\pi s}\frac{\G(\bar{\l}_2-s)}{\G(h-\D)}=\frac{-\pi}{\sin\pi\l_1}\frac{\G(1+\D-h)}{\G(1+s-\bar{\l}_2)}\,,
\ee
from which,
\begin{align}
\begin{split}
=\lim_{z\rightarrow0}G_{\D,J}(1-z,1-\bz)\bigg|_{\log z}=&-\frac{\G(2\l_1)\G(1+\D-h)}{\G(\l_1)^4(d-2)_J} \sum_{m=0}^J A_m(J,\D)\int_\mc ds\ \G(s)\bigg(\frac{1-\bz}{\bz}\bigg)^{s}\\
&\times\frac{\pi}{\sin\pi \l_1}\frac{\G(\l_1-m-s)\G(1+m+s-\l_1-\bar{\l}_2)}{\G(1-\l_1+m)\G(1+s-\bar{\l}_2)\G(1-\bar{\l}_2)}\,,
\end{split}
\end{align}
Finally we shift $s\rightarrow s+\l_1-m$ so that, 
\begin{shaded}
\begin{align}\label{llog}
\begin{split}
&G^t_{\D,J}|_{\log}=\lim_{z\rightarrow0}G_{\D,J}(1-z,1-\bz)\bigg|_{\log z}\\
=&-\frac{\G(2\l_1)}{\G(\l_1)^2(d-2)_J} \sum_{m=0}^J (-1)^m\frac{A_m(J,\D)}{(\l_1-m)_m^2}\int_\mc ds\ \bigg(\frac{1-\bz}{\bz}\bigg)^{\l_1-m+s} \frac{\G(-s)(\l_1-m)_s(1-\bar{\l}_2)_s}{(1+\D-h)_{s+J-m}}\,,
\end{split}
\end{align}
\end{shaded}
The $s=n$ poles (closing the contour along $\mc$) reproduces the physical block. 

\subsection{Regular term}
Similar to the $\log$, term we can evaluate the regular term (the correction to the OPE coefficients) by following the exact same steps as above. Starting with the second term in \eqref{logreg}, and performing the $k-$sum, 
\begin{align}\label{opechannel}
\begin{split}
&\lim_{z\rightarrow0}G_{\D,J}(1-z,1-\bz)\bigg|_{\text{ reg}}\\
=&-\frac{\G(2\l_1)\G(1+\D-h)}{\G(\l_1)^4(d-2)_J}\frac{\pi}{\sin\pi \l_1}\int_\mc ds\ \frac{\G(s)}{\G(1+s-\bar{\l}_2)\G(1-\bar{\l}_2)}\bigg(\frac{1-\bz}{\bz}\bigg)^{s}\\
&\times\bigg[\sum_{m=0}^J A_m(J,\D)\frac{\G(\l_1-m-s)\G(1+m+s-\l_1-\bar{\l}_2)}{\G(1-\l_1+m)}(H_{s-1}-\pi\cot\pi s+H_{m-\l_1}\\
&-H_{s-\l_1+m-\bar{\l}_2}+H_{-\bar{\l}_2}) +\sum_{1\leq m+n\leq J} B_{m,n}(J,\D)\frac{\G(\l_1-m-n-s)\G(1+m+n+s-\l_1-\bar{\l}_2)}{\G(1-\l_1+m+n)}\bigg]\,,
\end{split}
\end{align}
To get the final form we can shift the variables $s\rightarrow s+\l_1-m$. However, in order to keep coherence with the expressions used in the main text, we will treat the integrals separately, by writing,  
\begin{align}
\begin{split}
&\lim_{z\rightarrow0}G_{\D,J}(1-z,1-\bz)\bigg|_{\text{ reg}}\\
=&-\frac{\G(2\l_1)\G(1+\D-h)}{\G(\l_1)^4(d-2)_J}\frac{\pi}{\sin\pi \l_1}\bigg[\sum_{m=0}^J A_m(J,\D)\\
&\times\int_\mc ds\ \frac{\G(s)\G(\l_1-m-s)\G(1+m+s-\l_1-\bar{\l}_2)}{\G(1+s-\bar{\l}_2)\G(1-\bar{\l}_2)\G(1-\l_1+m)}\bigg(\frac{1-\bz}{\bz}\bigg)^{s}(H_{s-1}-\pi\cot\pi s+H_{m-\l_1}-H_{s-\l_1+m-\bar{\l}_2}\\&+H_{-\bar{\l}_2})+\sum_{1\leq m+n\leq J} B_{m,n}(J,\D)\int_\mc ds\ \frac{\G(s)\G(\l_1-m-n-s)\G(1+m+n+s-\l_1-\bar{\l}_2)}{\G(1+s-\bar{\l}_2)\G(1-\bar{\l}_2)\G(1-\l_1+m+n)}\bigg(\frac{1-\bz}{\bz}\bigg)^{s}\bigg]\,.
\end{split}
\end{align}
Note that the general contour $\mc$ works for both the integrals since we are choosing the contour in a way such that apart from the poles $s=\l_1-m+k$ for the first integral and $s=\l_1-m-n+k$ for the second integral, there are no new poles. For any general $m,n,k$ values the minimal pole in both the integrals is at $s=\l_2$ (for maximal $m$ and $m+n$ values and $k=0$). The contour $\mc$ is chosen such that $s=\l_2+n$ poles are always allowed. Now we shift $s\rightarrow s+\l_1-m$ in the first integral and $s\rightarrow s+\l_1-m-n$ in the second integral, so that, 
\begin{shaded}
\begin{align}
\begin{split}
&G^t_{\D,J}|_{\text{reg}}=\lim_{z\rightarrow0}G_{\D,J}(1-z,1-\bz)\bigg|_{\text{ reg}}\\
=&-\frac{\G(2\l_1)}{\G(\l_1)^2(d-2)_J}\bigg[\sum_{m=0}^J (-1)^m\frac{A_m(J,\D)}{(\l_1-m)_m^2}\int_\mc ds\ \frac{\G(-s)(\l_1-m)_s(1-\bar{\l}_2)_s}{(1+\l_2-\bar{\l}_2)_{s+J-m}}\bigg(\frac{1-\bz}{\bz}\bigg)^{s+\l_1-m}(H_{s+\l_1-m-1}\\
&-\pi\cot\pi (s+\l_1)+H_{m-\l_1}-H_{s-\bar{\l}_2}+H_{-\bar{\l}_2})\\
&+\sum_{1\leq m+n\leq J} (-1)^{m+n}\frac{B_{m,n}(J,\D)}{(\l_1-m-n)_{m+n}^2}\int_\mc ds\ \frac{\G(-s)(\l_1-m-n)_s(1-\bar{\l}_2)_s}{(1+\l_2-\bar{\l}_2)_{s+J-m-n}}\bigg(\frac{1-\bz}{\bz}\bigg)^{s+\l_1-m-n}\bigg]\,,
\end{split}
\end{align}
\end{shaded}

\noindent which is the starting point  for \eqref{mackterms} and \eqref{mackderv} in section \ref{opecorrec}.

\section{Integral involving harmonic number} \label{harmonicsumtrick}
In this appendix we will show in detail the steps required to evaluate the integral $I_1$ in (\ref{I1integral}). The difficulty in integrating it is that the expression contains a Harmonic number. Harmonic numbers are difficult to sum over once summing over residues. The equation (\ref{I1integral}) has the following functional form,
\begin{equation}\label{functionalform}
\int_\mc \: ds\, f(s,a,b,\cdots)\Gamma(s+k)H_{s+k-1} .
\end{equation}
We will solve this issue by generating this expression by differentiating one of the Gamma functions. We begin with the $I_1$ integral,
\begin{align}
\begin{split}\label{whatwewant}
I_1=\alpha_J \sum_{m=0}^J \frac{(-1)^{m+1}}{(\l_1-m)_m^2} \int_\mc &ds\G(-s) \frac{(\l_1-m)_s(1-\bar{\l}_2)_s}{(1+\l_2-\bar{\l}_2)_{s+J-m}}  \frac{\Gamma(\frac{\beta-\tau}{2}-\l_1+m -1 -s)\Gamma(\frac{\tau}{2}-m+\l_1 +1 +s)^2}{\Gamma(\frac{\beta+\tau}{2}-m +1 +s+\l_1)}\\
&\times A_m(J,\D)(H_{s+\l_1-m-1} -H_{s-\bar{\l}_2})\,.
\end{split}
\end{align}

To perform the integral we first take the integrand without the Harmonic numbers. We then shift the Gamma functions whose argument corresponds to the Harmonic numbers and shift them by an arbitrary variable $\epsilon$ and then perform the integral. In a functional form (\ref{functionalform}) this looks like,
\begin{equation}
\int_\mc \: ds \,f(n,a,\cdots)\Gamma(n+k+\epsilon)=g(k,a,\epsilon,\cdots) \, .
\end{equation}
Performing the shift and the integral gives rise to the following result for $I_1$, 
\begin{align}
\begin{split}\label{prederiv}
&\alpha_J \sum_{m=0}^J \frac{(-1)^{m+1}}{(\l_1-m)_m^2} A_m(J,\D) \int_\mc ds\G(-s) \frac{\G(s+\l_1-m+\epsilon) \G(s+1-\bar{\l}_2-\epsilon)}{\G(1+\l_2-\bar{\l}_2+s+J-m)}\\ & \frac{\G(1+\l_2-\bar{\l}_2)}{\G(\l_1-m) \G(1-\bar{\l}_2)}   \frac{\Gamma(\frac{\beta-\Delta-J-\tau}{2}+m -1 -s)\Gamma(\frac{\Delta+J+\tau}{2}-m +1 +s)^2}{\Gamma(\frac{\beta+\Delta+J+\tau}{2}-m +1 +s)} \,\\
=&\alpha_J \sum_{m=0}^J \frac{(-1)^{m+1}}{(\l_1-m)_m^2} A_m(J,\D) \frac{\G(1+\l_2-\bar{\l}_2)}{\G(\l_1-m) \G(1-\bar{\l}_2)}\\
&\times\Bigg( \frac{\Gamma \left(\l_1 - h+1-\epsilon \right) \Gamma \left(\l_1 - m+\epsilon\right)
	\Gamma \left(\frac{\tau}{2}+\l_1 - m+1\right)^2 \Gamma \left(m-1+\frac{\beta
		-\tau }{2}-\l_1 \right)}{\Gamma
	(-h+2\l_1-m+ +1) \Gamma \left(\frac{\beta  +\tau }{2}+\l_1 -m+1\right)} \\ &\, \times _4F_3\bigg[\begin{matrix}1-h+\l_1-\epsilon,\l_1-m+\epsilon,\frac{\tau}{2}+\l_1-m+1,\frac{\tau}{2}+\l_1-m+1\\ -h+2\l_1-m+1,\frac{-\beta+\tau}{2}+\l_1-m+2,\frac{\beta+\tau}{2}+\l_1-m+1\end{matrix};1\bigg]   \\
&+\frac{\Gamma \left(\frac{\beta }{2}\right)^2 \Gamma \left(\frac{\beta -\tau }{2} -1+\epsilon\right) \Gamma
	\left(- h+ m+\frac{\beta -\tau }{2}-\epsilon )\right) \Gamma \left(\frac{-\beta +\tau
	}{2}+\l_1 -m+1\right)}{\Gamma (\beta ) \Gamma \left(\frac{\beta -\tau }{2}+\l_1 -h\right)}  \\
&\times _4F_3\bigg[\begin{matrix}\frac{\beta }{2},\frac{\beta }{2},\frac{\beta -\tau}{2}-1+\epsilon,\frac{\beta -\tau}{2}-h+m-\epsilon\\ \beta,\frac{\beta-\tau }{2}-\l_1+m,-h+\frac{\beta  - \tau }{2}+\l_1\end{matrix};1\bigg] \Bigg).
\end{split}
\end{align}
Now we can take derivatives of both sides with respect to $\epsilon$ \footnote{This procedure is well defined if the infinite sum is convergent.} and set $\epsilon$ to 0. The left side becomes exactly the integral we wanted (\ref{whatwewant}) as derivative of Gamma functions generate Polygamma functions,
\begin{equation}
\frac{d}{d\epsilon} \Gamma(a+\epsilon)=\Gamma (a+\epsilon ) \psi ^{(0)}(a+\epsilon )=\Gamma (a+\epsilon ) (H_{a+\epsilon-1}-\gamma) \, .
\end{equation}
In the above equation $\gamma$ is the Euler-Mascheroni constant. The right side (\ref{prederiv}) result would involve derivative of ${}_4F_3$ with respect to its parameters. We will first discuss the expressions for derivative of hypergeometric functions in terms of Kamp\'e de F\'eriet-like functions. We begin with the expression of a generalized hypergeometric function \footnote{We follow the conventions and notations of \cite{hypderiv}.}\footnote{We will refer to $z$ as the argument of the hypergeometric.},
\begin{equation}\label{genhyp}
{}_pF_q(a_1,\ldots,a_p;b_1,\ldots,b_q;z) = \sum_{n=0}^\infty \frac{(a_1)_n\cdots(a_p)_n}{(b_1)_n\cdots(b_q)_n} \, \frac {z^n} {n!}=\sum_{n=0}^{\infty}  A_n\frac {z^n} {n!} \,.
\end{equation}
The parameters of each row of a hypergeometric are symmetric so we can just consider derivative of hypergeometric with respect to the first parameter $a_1$,
\begin{equation}
G_{a_1}=\frac{d\:( {}_pF_q[(a_1,\cdots,a_p );(b_1,\cdots,b_q );1])}{da_1} \,.
\end{equation}
Before proceeding further let us introduce some notations for Kamp\'e de F\'eriet-like double sum functions,
\begin{align}
&\pTTq{6}{5}{\alpha_1,\alpha_2\mid \,a_1,a_2,\cdots,a_p}{\gamma \mid b_1,b_2,\cdots,b_q}{x}{y} \\ \\ &\qquad \qquad= \sum_{m=0}^{\infty}\sum_{n=0}^{\infty}\frac{(\alpha_1)_m(\alpha_2)_n(a_1)_m}{(\gamma)_m}\frac{(a_2)_{m+n}(\cdots)(a_p)_{m+n}}{(b_1)_{m+n}(b_2)_{m+n}(\cdots)(b_q)_{m+n}}\frac{x^m y^n}{m!n!} \, .
\end{align}
In this paper we only encounter the derivatives of ${}_4F_3$ with argument 1. Specializing the above equation to our case we obtain,
\begin{equation}
\frac{d ({}_4F_3(1))}{d a_1}=G_{a_1}(1)=\frac{1}{a_1}A_1\: \pTq{6}{5}{1,1\mid\,a_1,a_1+1,a_2+1,a_3+1,a_4+1}{a_1+1\mid 2,b_1+1,b_2+1,b_3+1} \, .
\end{equation}
The hypergeometric and its derivatives have argument 1 and $A_1$ is defined in (\ref{genhyp}). For convenience we provide the expanded expression for the double sum below,
\begin{align}
&\pTq{6}{5}{1,1\mid\,a_1,a_1+1,a_2+1,a_3+1,a_4+1}{a_1+1\mid 2,b_1,b_2,b_3}\\ \\
&\hspace{4 cm}= \sum_{m=0}^{\infty}\sum_{n=0}^{\infty}\frac{(a_1)_m}{(a_1)_{m+1}}\frac{(a_1+1)_{m+n}(a_2+1)_{m+n}(a_3+1)_{m+n}(a_4+1)_{m+n}}{(2)_{m+n}(b_1)_{m+n}(b_2)_{m+n}(b_3)_{m+n}} \, .
\end{align}
We will further define two notations which will be used in the maintext,
\begin{align}\label{kampeG}
\mathcal{G}_1 \,=&\frac{A_1}{a_1}\: \pTq{6}{5}{1,1\mid\,a_1,a_1+1,a_2+1,a_3+1,a_4+1}{a_1+1\mid 2,b_1+1,b_2+1,b_3+1}{}{} - (a_1 \leftrightarrow a_2)\, , \nonumber \\ 
\mathcal{G}_2 \,=&\frac{C_1}{c_1}\: \pTq{6}{5}{1,1\mid\,c_1,c_1+1,c_2+1,c_3+1,c_4+1}{c_1+1\mid 2,d_1+1,d_2+1,d_3+1}{}{} - (c_1 \leftrightarrow c_2)\, , 
\end{align}		
with ,
\begin{align}
\begin{split}
a_1 &= 1-h+\l_1, \hspace{2.2cm} a_2 = \l_1 -m, \hspace{2.3cm} a_3 = a_4 =\frac{\tau}{2}+\l_1-m+1, \\ b_2 &= \frac{\tau-\beta }{2}+\l_1-m+2, \hspace{0.65cm} b_1 = 1-h+J-m,  \hspace{1.15cm} b_3= \frac{\tau-\beta }{2}+\l_1-m+1, \\
c_1 &= \frac{\beta-\tau}{2}-1, \hspace{2.3cm} c_2 = \frac{\beta-\tau}{2} - h + m, \hspace{1.05cm}
c_3 = c_4 = \frac{\beta}{2}, \\ 
d_2 &= \frac{\beta-\tau}{2}-\l_1+m, \hspace{1.3cm} d_1 = \beta, \hspace{3.3cm} d_3 = \frac{\beta-\tau}{2}+\l_1-h, \\ A_1&= \frac{\prod_{i=1}^4\,a_i}{\prod_{i=1}^3\,b_i}, \hspace{2.6cm} C_1=\frac{\prod_{i=1}^4\,c_i}{\prod_{i=1}^3\,d_i}.
\end{split}
\end{align}
Returning to (\ref{prederiv}), on taking the derivative of the right side with $\epsilon$ and using the expression for double sums we obtain,
\begin{align}
\begin{split}
&\alpha_J \sum_{m=0}^J \frac{(-1)^{m+1}}{(\l_1-m)_m^2} \int_\mc ds\G(-s) \frac{(\l_1-m)_s(1-\bar{\l}_2)_s}{(1+\l_2-\bar{\l}_2)_{s+J-m}}  \frac{\Gamma(\frac{\beta-\tau}{2}-\l_1+m -1 -s)\Gamma(\frac{\tau}{2}-m+\l_1 +1 +s)^2}{\Gamma(\frac{\beta+\tau}{2}-m +1 +s+\l_1)}\\
&\times A_m(J,\D)\,(H_{s+\l_1-m-1} -H_{s-\bar{\l}_2}) \\
=&\alpha_J \sum_{m=0}^J \frac{(-1)^{m+1}}{(\l_1-m)_m^2} A_m(J,\D) \frac{\G(1+\l_2-\bar{\l}_2)}{\G(\l_1-m) \G(1-\bar{\l}_2)}\\
\Bigg[& C_m(J,\Delta) \Bigg( \,\,\, _4F_3\bigg[\begin{matrix}1-h+\l_1,\l_1-m,\frac{\tau}{2}+\l_1-m+1,\frac{\tau}{2}+\l_1-m+1\\ -h+2\l_1-m+1,\frac{-\beta+\tau}{2}+\l_1-m+2,\frac{\beta+\tau}{2}+\l_1-m+1\end{matrix};1\bigg]    \\ &( H_{\l_1-m-1} - H_{-h+\l_1} ) -\mathcal{G}_1\,\Bigg) \\
&+D_m(J,\Delta)\Bigg(\,\,\,_4F_3\bigg[\begin{matrix}\frac{\beta }{2},\frac{\beta }{2},\frac{\beta -\tau}{2}-1,\frac{\beta -\tau}{2}-h+m\\ \beta,\frac{\beta-\tau }{2}-\l_1+m,-h+\frac{\beta  - \tau }{2}+\l_1\end{matrix};1\bigg] (H_{\frac{\beta -\tau}{2}-2} - H_{\frac{\beta -\tau}{2}-h+m-1})+\mathcal{G}_2\Bigg)\Bigg].
\end{split}
\end{align}
This completes our derivation of (\ref{I1spinj}).

\section{Wilson function}

We now describe the procedure to write the two $_4 F_3$s as a single $_7 F_6$. Following the conventions of  \cite{Hogervorst:2017sfd}, Wilson function can be written as a linear combination of two balanced $_4F_3(1)$ as,  
\begin{align}
\phi_\alpha(\beta;a,b,c,d) = \frac{\G(d-a)}{\G(a+b)\G(a+c)\G(d\pm\beta)\G(\tilde{d}\pm\alpha)}\, _4F_3\bigg[\begin{matrix}a+\beta,a-\beta,\tilde{a}+\alpha,\tilde{a}-\alpha\\ a+b,a+c,1+a-d\end{matrix};1\bigg]    +(a\leftrightarrow d), 
\end{align}
where, 
\begin{align}
\begin{split}
\tilde{a} =& \frac{1}{2}(a+b+c-d), \hspace{2cm} \tilde{d} = \frac{1}{2}(-a+b+c+d), \\ 
\tilde{b} =& \frac{1}{2}(a+b-c+d), \hspace{2cm} \tilde{c} = \frac{1}{2}(a-b+c+d),\\
&\qquad \qquad \Gamma(a\pm b)=\Gamma(a+b)\Gamma(a-b).
\end{split}    
\end{align}
Wilson function can also be written \cite{2003math......6424G}\footnote{Our conventions are related to those of \cite{2003math......6424G} as $\phi_\alpha(\beta;a,b,c,d)=\phi_{i\, \alpha}(i\,\beta;a,b,c,1-d)$.}  as,
\begin{align}
\phi_\alpha(\beta;a.b.c.d) = &\frac{\Gamma (\tilde{a}+\tilde{b}+\tilde{c}-\alpha )}{\Gamma (a+b) \Gamma (a+c) \Gamma (a+d) \Gamma
   (\tilde{d}-\alpha ) \Gamma (\tilde{b}+c-\alpha -\beta) \Gamma (\tilde{b}+c-\alpha +\beta)}\nonumber \\
&\times W(\tilde{a} +\tilde{b}+\tilde{c}-1-\alpha;a-\beta,a+\beta,\tilde{a}-\alpha,\tilde{b}-\alpha,\tilde{c}-\alpha),  
\end{align}
where the $W$-function above can be written as a ${}_7F_6$,
\begin{align}
W(a;b,c,d,e,f)= \,_7F_6\bigg[\begin{matrix}a\,,\frac{a}{2}+1\,,b\,,c\,,d\,,e\,,f\\\frac{a}{2},1+a-b,1+a-c,1+a-d,1+a-e,1+a-f\end{matrix};1\bigg].   
\end{align}
This procedure gives us same final result as the one given in \cite{Gopakumar:2018xqi}.

\bibliographystyle{JHEP}
\bibliography{LSSCH}
\end{document}